\documentclass{article}

\usepackage[a4paper, margin=2cm]{geometry}

\usepackage[T1]{fontenc}
\usepackage{lmodern}
\usepackage{microtype}
\usepackage{listings}
\usepackage{xcolor}

\usepackage{graphicx}
\usepackage[english]{babel}
\usepackage{amsmath}
\usepackage{mathtools}
\usepackage{bm}
\usepackage{comment}
\usepackage{hyperref}
\usepackage[super,sort&compress]{natbib}

\usepackage{authblk}
\usepackage{booktabs}
\usepackage{longtable}
\usepackage{siunitx}
\usepackage{array}

\usepackage[font=small,labelfont=bf]{caption}
\usepackage{subcaption}

\usepackage{parskip}

\title{Real-world and simulated thermal data from 960 residential multi-zone buildings in Central Europe}

\author[1,2,*]{Fabian Raisch}
\author[3]{Matthias Kersken}
\author[4]{Markus Male}
\author[1]{Benjamin Tischler}

\affil[1]{Technical University of Applied Sciences Rosenheim, Germany}
\affil[2]{Technical University of Munich, Germany}
\affil[3]{Fraunhofer Institute for Building Physics IBP, Germany}
\affil[4]{iDM Energiesysteme GmbH, Austria}
\affil[*]{Corresponding author: \texttt{fabian.raisch@th-rosenheim.de}}

\date{}
\begin{document}
\maketitle

\section*{Abstract}

This paper presents the ThermBuild dataset, which comprises real-world measurements from two single-family homes and simulations of 958 TRNSYS building models. The buildings cover diverse combinations of air-source heat pump systems, numbers of thermal zones, occupancy profiles, building ages, thermal masses, sizes, orientations, window glazing types, European climates, and ventilation configurations. The dataset contains 15-minute-resolution operational data spanning 15 months for the real-world buildings and 3 years for the simulated buildings. Each building time series includes detailed measurements of heat pump operation, the heating distribution system, the domestic hot water system, zone-level indoor climate variables, and weather conditions. \\
The ThermBuild dataset is designed for data-driven thermal modeling of buildings, thereby supporting the deployment of energy-efficient control, as well as fault detection and diagnosis. It is particularly suited for transfer learning, generalization modeling, benchmarking, simulation-to-reality transfer, and reproducible thermal modeling research.

\section{Background \& Summary}
\label{sec:intro}

Building operations account for approximately 30\% of global energy consumption and 26\% of global greenhouse gas emissions \cite{eea2023DecarbonisingHeatingCooling}. The building sector is therefore widely recognized as a key domain for reducing emissions to achieve climate targets. At the same time, building operations offer significant opportunities for emission reduction. Strategies such as automated fault detection and diagnosis (FDD) \cite{CHEN2023121030} and energy-efficient control approaches via model predictive control (MPC) \cite{Drgona.2020} and reinforcement learning (RL) \cite{NAGY2023110435} have demonstrated the potential to significantly reduce building energy usage, related emissions, and operational costs.
The deployment of these methods mostly requires accurate models of building thermal dynamics, including the heating, ventilation, and air-conditioning (HVAC) system. One approach is the use of manually calibrated white-box models. However, their development requires expert knowledge, extensive engineering effort, and significant calibration time \cite{Drgona.2020}. Data-driven gray-box and black-box models provide a more scalable alternative because they can be generated directly from measurement data. 
Despite these opportunities, most buildings worldwide are not equipped with sensors, limiting the potential for advanced operational methods. Accordingly, when FDD or advanced control systems are deployed in practice, historical data for the specific target buildings are often unavailable, which delays model generation and algorithm deployment. Publicly available datasets can help address this challenge by enabling the pretraining of models and policies prior to deployment, as demonstrated in recent studies \cite{raisch2025gentlgeneraltransferlearning, RAISCH2026116868, koch2026thermalgems, li2024building, dou2025transfer, devargas2026, raisch2026transferlearningneuralparameter}. Once trained, these models can be adapted to unseen target buildings, reducing implementation time and improving early-stage performance, a process well known as transfer learning. Effective pretraining and robust model evaluation require datasets that span a wide range of building characteristics, HVAC configurations, occupant behaviors, and climatic conditions \cite{raisch2025gentlgeneraltransferlearning, pinto2022sharing, li2024building}.
Access to datasets spanning diverse building types is, therefore, essential for transfer learning. 
Yet, most transfer learning studies in the building domain have evaluated their methods only on synthetic single-zone buildings. Publicly available datasets that combine simulated and real-world measurements of multi-zone buildings are therefore important for investigating the applicability of transfer learning methods under real operating conditions and for analyzing the simulation-to-reality transfer performance.
In addition, publicly available datasets support reproducible research and benchmarking for transfer learning and other data-driven methods. 

\newpage
For these reasons, we developed the \textbf{ThermBuild} dataset, which complements measurements from two real-world buildings with data from 958 simulated buildings. The real-world data originates from a 15-month measurement campaign conducted in the two TwinHouse reality labs in Holzkirchen, Germany, as used in \cite{Strachan_Svehla_Heusler_Kersken_2016, Kersken_Rojas_Strachan_2025}.
The simulated data is generated from a validated TRNSYS simulation environment and provides 3 years of data for each building. The entire dataset represents multi-zone single-family houses located in Central Europe, constructed between 1980 and today. The buildings cover variations in weather, occupancy, building properties, and two heat pump configurations. Building properties are represented by different building ages, thermal masses, window glazing types, orientations, sizes, and numbers of zones. Occupancy affects internal heat gains, domestic hot water (DHW) draw-offs, and different air-changing rates via natural or mechanical ventilation. Both simulated and measured data are provided at 15-minute resolution and cover heating and cooling periods. Each time series includes room-level indoor climate measurements, weather conditions, and detailed operational data from the heat pump, the distribution system including thermal storage, and DHW. 
The dataset contains all variables required to model the thermal dynamics of both the building envelope and the heating/cooling system. These thermal models can subsequently be used for energy-efficient heat pump and zone-level control, as well as fault detection and diagnosis. The dataset is designed to promote research on multi-zone thermal modeling, transfer learning, model and policy generalization, and the simulation-to-reality transfer challenge. For these domains, the dataset supports benchmarking and reproducibility research. Beyond thermal modeling, the dataset can be used for a broad range of related applications, including electrical and thermal load forecasting, heat pump performance analysis, and thermal energy storage modeling. 

Several other building datasets exist, such as the Building Data Genome Project \cite{miller2020buildinggenome2} and BuildingBench \cite{emami2023buildingsbench}, which provide large-scale electricity load data with associated metadata. These datasets focus primarily on the electrical side of the building and provide limited insight into thermal dynamics and HVAC behavior.
Few datasets target the thermal side of residential buildings. The ecobee dataset \cite{luo2022ecobee} provides thermostat data from around 1,000 homes but lacks detailed HVAC measurements and multi-zone observations. The HOT dataset \cite{berkes2025hot} scales to approximately 150,000 synthetic homes, yet does not represent real-world operation or detailed HVAC time series. The IDEAL household energy dataset \cite{pullinger2021ideal} includes electricity, gas, and multi-zone sensing for 255 homes, offering richer heating-related context, although the time series have inconsistent measurements (only 32 out of the 255 homes include indoor temperatures and radiator flow temperatures), weather data is missing, and no heat pump systems are included. 
Open-source data generators provide an alternative approach for acquiring thermal building data \cite{eplusr, chaudhary2023synconn_build}. For example, \cite{11118811, krug2025builda2, koch2026buildynexcitationdrivendatageneration} offers a Modelica-based framework with a Python interface that enables the automated configuration and simulation of multiple buildings. However, the framework is limited to single-zone buildings and represents HVAC systems only in a simplified manner.

\section{Methods}
\label{ch:methods}

The ThermBuild dataset consists of two complementary parts. The first part, further described in Section~\ref{ch:RealityLab}, contains real-world measurement data collected from two residential buildings equipped with different air-source heat pump systems. The second part, further described in Section~\ref{ch:Simulation study}, contains simulated data generated using the TRNSYS 18 thermal-energetic building simulation software \cite{Klein2017TRNSYS18}. This component provides three years of time series data for 958 simulated buildings derived from variations of the real-world buildings.

\subsection{Real-world measurements}
\label{ch:RealityLab}

\begin{figure}[tbp] 
\centering 
\begin{subfigure}[t]{0.49\textwidth} 
\centering 
\includegraphics[width=\linewidth]{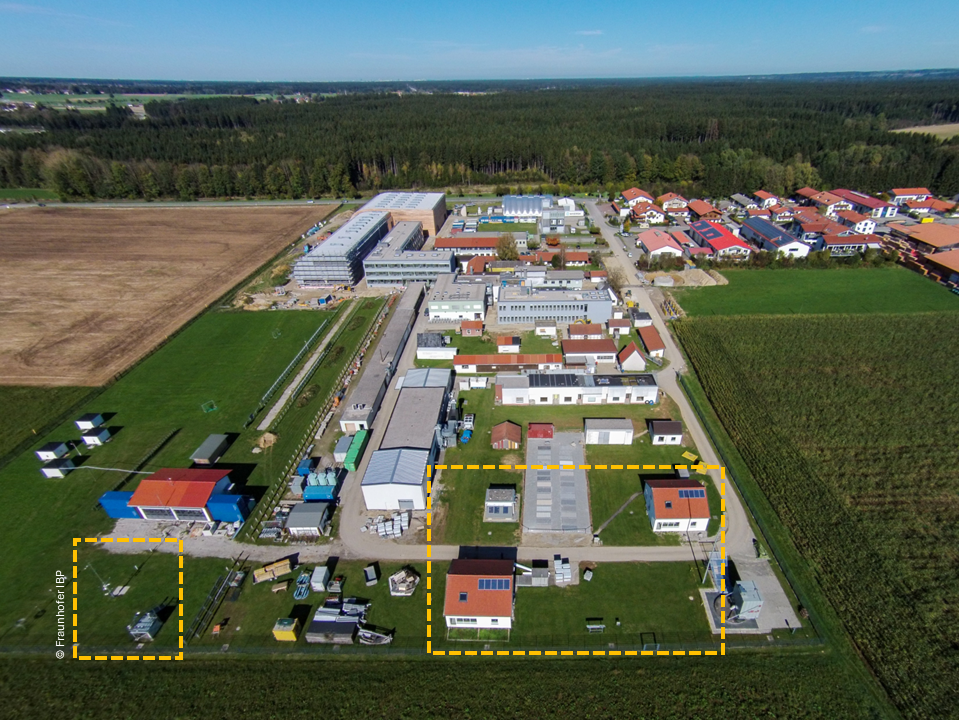} 
\subcaption{Aerial view from the south of the two TwinHouses (right) and the weather station (left) marked in yellow.} \label{fig:ZHout_left} 
\end{subfigure} 
\begin{subfigure}[t]{0.49\linewidth} 
\centering \includegraphics[width=\linewidth]{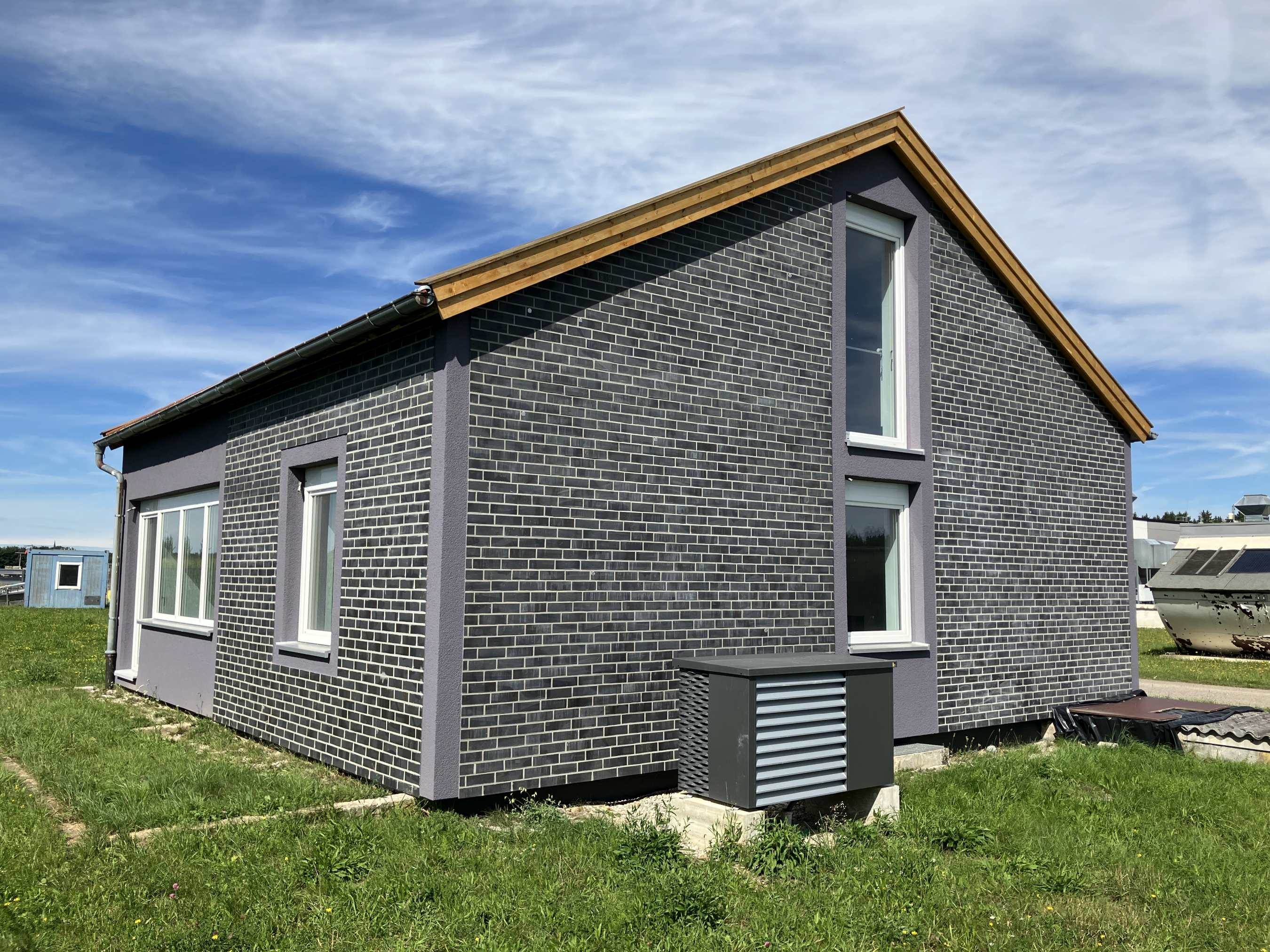} 
\subcaption{West view of the N2 TwinHouse, including the outside unit of the heat pump.} 
\label{fig:ZHout_right} 
\end{subfigure} 
\caption{Fraunhofer IBP TwinHouse test site near Holzkirchen (Germany) and installed heat pump system.} 
\label{fig:ZHout} 
\end{figure}

The real-world measurements were conducted from 7 February 2025 to 30 April 2026 in two residential test buildings, referred to as the TwinHouses. The TwinHouses have previously been used in several scientific studies, including the validation of building energy simulation tools within the framework of IEA EBC Annex 58 \cite{Strachan_Svehla_Heusler_Kersken_2016} and Annex 71 \cite{Kersken_Rojas_Strachan_2025}. Figure~\ref{fig:ZHout_left} shows the test site including the two TwinHouses and the institute's weather station. The eastern house is called TwinHouse N2 (see Figure \ref{fig:ZHout_right}), and the western TwinHouse O5. The buildings have been set up as two identical detached single-family houses at the Fraunhofer Institute for Building Physics IBP near Holzkirchen, south of Munich, Germany. Each building comprises a living space of 140 m², which corresponds to the German average for single-family homes. The TwinHouses were constructed around 1980 and have been retrofitted continuously to meet modern thermal requirements. Currently, the thermal status meets the requirements for a newly refurbished building according to the German building energy act (GEG~2020) \cite{GEG2024}. Both houses are equipped with standard triple-pane insulating glazing and comprise seven common rooms. The basements were excluded from the investigation and maintained at a constant temperature of 18 \textdegree C by a separate heating system. The buildings experience only negligible external shading and share the same orientation, with the living room facing south.

\begin{figure}[t]
    \centering
    \vspace*{-0.5cm}
    \includegraphics[width=0.8\linewidth]{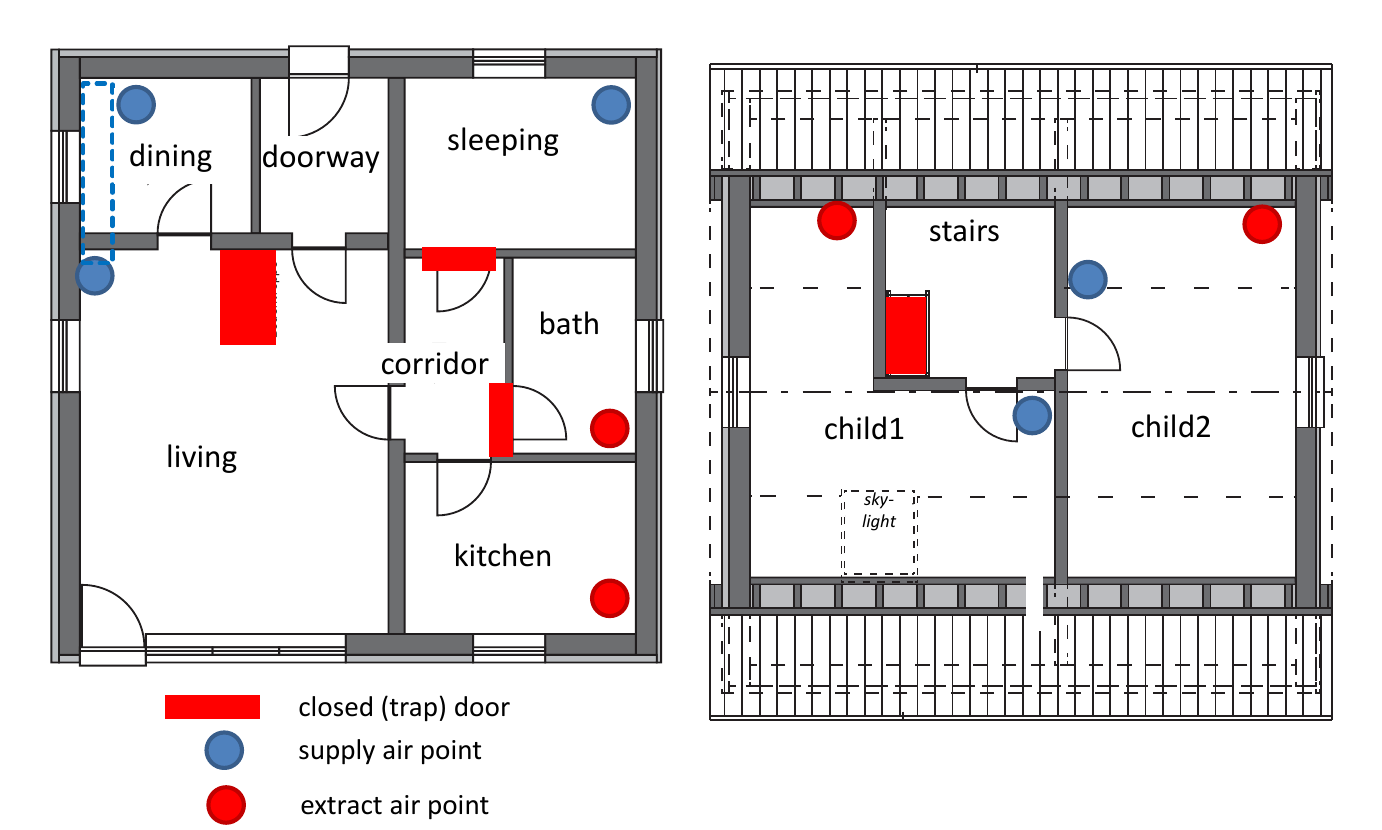}
    \vspace*{-0.2cm}
    \caption{Floor plan of the ground floor (left) and the attic (right) of the TwinHouses, including the indication of open and closed doors as well as the supply and extract air positions for the mechanical ventilation system.}
    \label{fig:TH_FloorPlan}
\end{figure}

Figure~\ref{fig:TH_FloorPlan} provides the TwinHouses' floor plan and the setup of the mechanical ventilation system. The doors from the corridor to the sleeping room and to the bathroom are closed, as is the trapdoor between the ground floor and the attic. Figure~\ref{fig:TH_FloorPlan} also shows the supply and extract air points of the mechanical ventilation system. Dining, sleeping, and the living room are supply air rooms, while the kitchen and bath have extract air points. The two child rooms in the attic have balanced supply and extract air. 

\begin{figure}[tb]
    \centering
    \begin{subfigure}[t]{0.49\linewidth}
        \centering
        \includegraphics[height=5cm]{figures/TGA_Keller_O5_iPump.JPG}
        \caption{BSE1: iDM iPump compact unit in the O5 basement, including a 50~L heating buffer for the defrost cycles (integrated DHW buffer was not used).}
        \label{fig:ZH_TGA_keller_left}
    \end{subfigure}
    \hfill
    \begin{subfigure}[t]{0.49\linewidth}
        \centering
        \includegraphics[height=5cm]{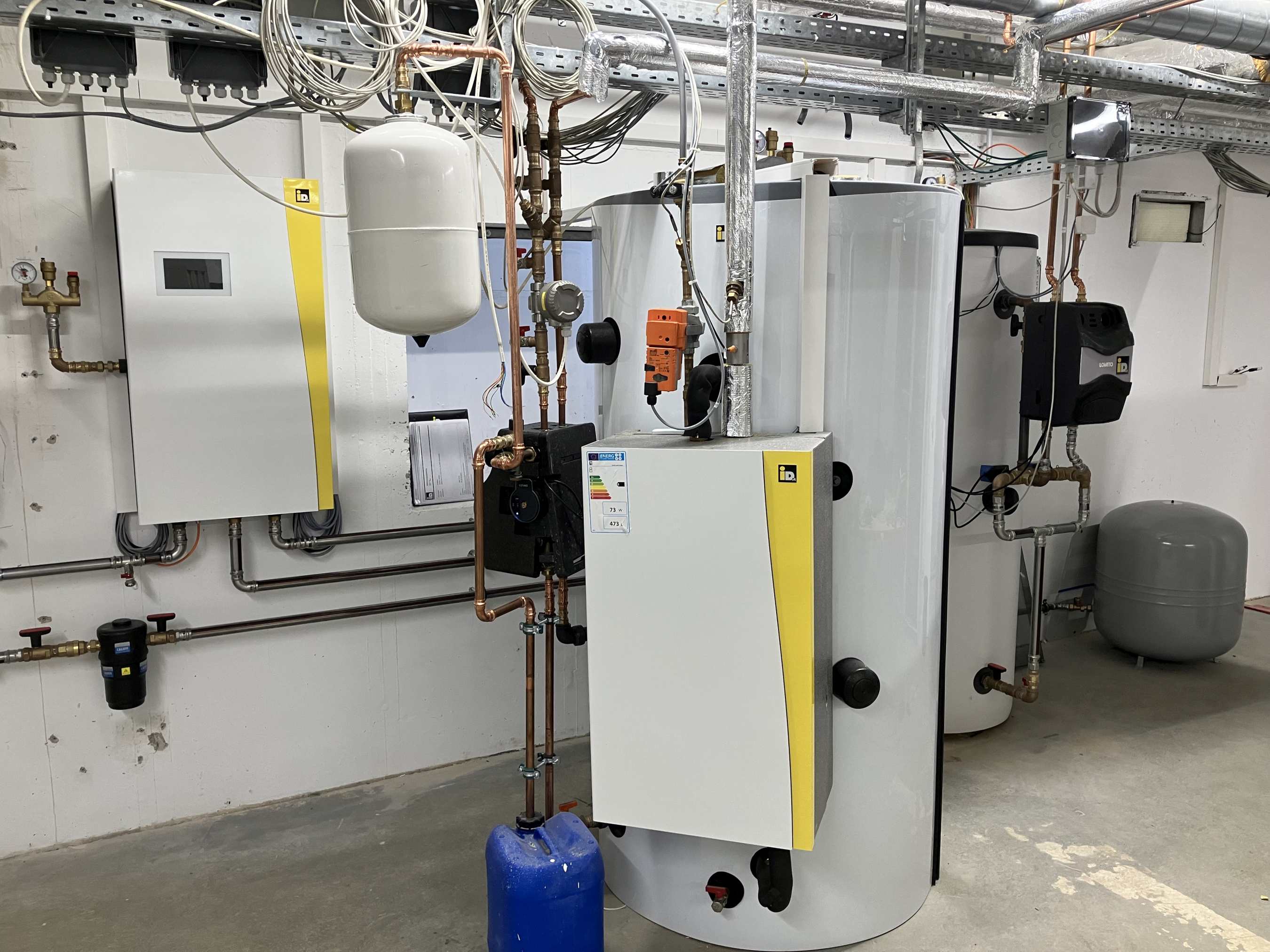}
        \caption{BSE2: iDM Alm consisting of wall-mounted indoor unit, DHW solar storage tank, including fresh water station and heating buffer (left to right) in the N2 basement.}
        \label{fig:ZH_TGA_keller_right}
    \end{subfigure}
    \caption{Building service equipment installation.}
    \label{fig:ZH_TGA_keller}
\end{figure}

The principal distinction between the two TwinHouses lies in their building service equipment (BSE) configurations. TwinHouse O5 is equipped with a compact heat pump system featuring only minimal thermal storage capacity, which is primarily required for heat pump defrost operation. The system, therefore, supplies the building almost directly without thermal buffering in between. This configuration is referred to as \textbf{BSE1} and is illustrated in Figure~\ref{fig:ZH_TGA_keller_left}. BSE1 does not include DHW production.
TwinHouse N2 is equipped with a heat pump system comprising a wall-mounted indoor unit, an 825~L DHW storage, and a 500~L heating buffer. DHW production is additionally supported by a solar thermal collector with an area of approximately 6 m². DHW is drawn from the storage tank according to a predefined occupancy-based tapping profile. This configuration is referred to as \textbf{BSE2} and is illustrated in Figure~\ref{fig:ZH_TGA_keller_right}. Together, BSE1 and BSE2 represent direct-coupled and storage-coupled heat pump systems, resulting in distinct thermal dynamics.
Both TwinHouses use identical outdoor units for the air-source heat pumps, as shown for TwinHouse N2 in Figure~\ref{fig:ZHout_right}. In both TwinHouses, heating and cooling are distributed through underfloor heating systems, consisting of wet-screed underfloor heating on the ground floor and dry-screed underfloor heating in the attic.

Occupancy is represented through internal heat gains and the definition of thermal comfort conditions. For internal gains, the TwinHouses include room-wise electrical power-controlled heat gain simulators with a convective-radiative split of approximately 50\%. For the occupancy profiles, we used the Occdem software \cite{Flett2021OccDem}. Further information on software usage and occupancy will be provided in Section~\ref{sec:occupancy}. The setpoint temperatures in both houses are set to 20~\textdegree C for all rooms with no night setback. The buildings are equipped with a mechanical ventilation system operating at a constant building-wide airflow rate of 155~m\textsuperscript{3}/h, as specified by the German building energy act for the calculation of mechanical ventilation systems \cite{GEG2024}.

The data acquisition systems consisted of the Beckhoff PLC of the TwinHouses with a 1-second sampling interval and the internal logging system of the iDM heat pumps. When measurements were available from both systems, the TwinHouses data were preferred because the instrumentation was calibrated prior to the experiment. Due to various technical issues, the raw data contains gaps in one or both systems. When data were available from the complementary iDM acquisition system, the missing values were replaced accordingly. If gaps persisted in both systems, a k-nearest neighbor (kNN) method with a neighborhood size of 5 was applied for imputation. The datasets before and after imputation are provided separately. Additional details on the kNN implementation are presented in Section~\ref{sec:code}.

\subsection{Simulation study}
\label{ch:Simulation study}

\begin{figure}[tbp]
    \centering
    \includegraphics[width=0.7\linewidth]{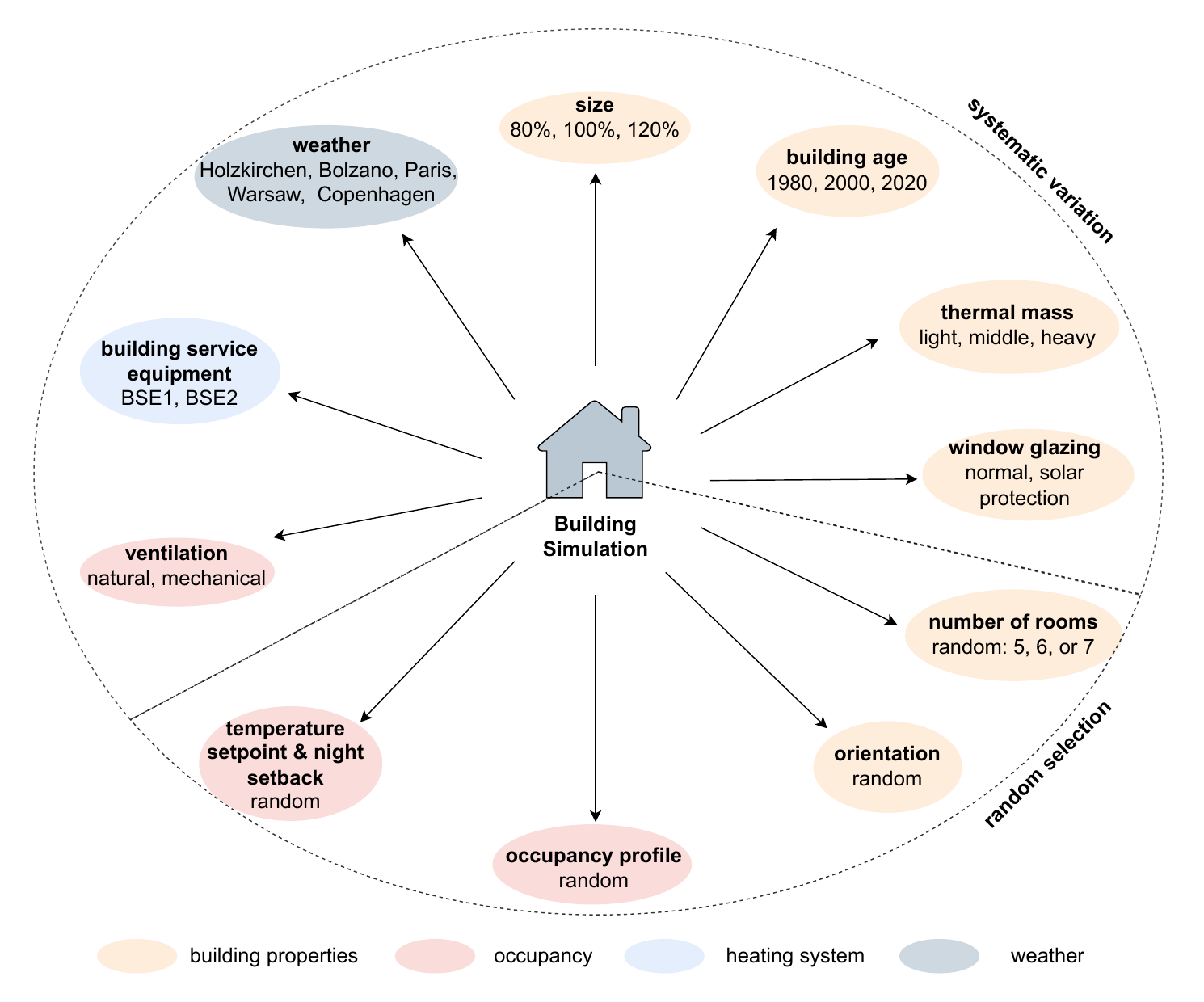}
    \caption{Overview of the simulation study variations. A total of 1,080 systematic building configurations were generated, while parameters marked as random selection were assigned individually for each simulation case.}
    \label{fig:fig_variations}
\end{figure}

The aim of the simulation study is to complement the TwinHouses measurement data with synthetic data. To this end, the TwinHouses were modeled in the TRNSYS simulation environment to establish two baseline models. Variations of these two baselines were generated across three domains: building properties, occupancy, and weather conditions. Each domain comprises multiple systematically varied sub-parameters, as illustrated in Figure~\ref{fig:fig_variations}. These parameters were selected because they most strongly influence the thermal dynamic behavior of buildings \cite{Thomas2006Environmental}. Sections~\ref{sec:building_prop} to \ref{sec:weather} describe all parameters and variations in detail. As illustrated in Figure~\ref{fig:fig_variations}, the parameter variations were divided into systematic variations and random selection. Systematic variations were generated by defining a discrete set of values for each parameter and combining all values across variations. These variations comprise 2 ventilation types, 5 locations, 3 building sizes, 3 building ages, 3 thermal mass categories, 2 window glazing types, and the 2 BSE configurations derived from the baseline TwinHouses. 
Combining all systematic variations results in 1,080 unique building configurations. In addition, each configuration was assigned a set of randomly selected parameters, including temperature setpoints, occupancy profiles, building orientation, and the number of rooms.
Based on these configurations, 1,080 TRNSYS models were automatically generated from the two baselines. Due to the automated model generation process, some simulation cases failed to meet the predefined convergence criteria and were therefore discarded.
The convergence tolerance was set to 0.1\%, defining the maximum allowed deviation between successive iteration steps for each 5-minute simulation timestep. Up to 400 iterations were permitted per timestep, and simulations were discarded if more than 300 timesteps (0.3\%) failed to converge.
The final dataset consequently contains 958 successful simulation runs with a total computation time of about 8.5 days.

The simulated dataset was generated at a 15-minute temporal resolution to maintain consistency with the real-world measurements. To improve numerical stability, the TRNSYS simulations were performed with a 5-minute timestep and subsequently aggregated to 15-minute mean values. 

In the following, all parameter variations of the simulation study are explained.


\subsubsection{Building properties}
\label{sec:building_prop}

The varied building properties include building size, building age, thermal mass, window glazing, number of rooms, and orientation.

\paragraph{Size} The size of the building influences the amount of energy required for heating. To capture the influence of building size, the floor area of the baseline models is scaled to 80\% and 120\%. The TwinHouses' living space area (140 m²) is defined as 100\% because it represents the average German single-family house. The surface areas of the walls, roof, windows, and internal components (incl. underfloor heating) are adjusted accordingly.

\paragraph{Building age}
The thermal performance of a building is strongly influenced by the thermal resistance of its envelope, which is closely related to the construction period and prevailing building standards at the time of construction \cite{tabula}. To represent this variation in the simulation study, three different building ages are chosen. Besides the current energetic quality of the TwinHouses, year 2020, building ages according to the year 1980 and 2000 are chosen. Table~\ref{tab:tab_envelope_combined} lists the U-values of the envelope's main components according to the chosen ages. The 2020 U-values are defined by TwinHouses' current constructions. For the other two construction ages, these properties are taken from the ``announcement of the rules for data collection and data use in existing residential buildings'' of the German federal ministries for economic affairs and energy, and the federal ministry of the interior, building, and community \cite{bbsr_datenaufnahme_wohngeb_geg}. The U-values are provided in Table~\ref {tab:tab_envelope_combined} for each building age. 

\begin{table}[b]
\centering
\caption{Envelope U-values and glazing g-values of the TwinHouses (2020) and the simulation study for different building ages. The envelope mean was calculated based on the area-weighted U-values of all envelope components.}
\label{tab:tab_envelope_combined}
\begin{tabular}{c|c|cccc|cc}
 &  \multicolumn{5}{c|}{U-value [W/(m²K)]} & \multicolumn{2}{c}{g-value [-] of glazing} \\ 
age 
& envelope mean
& external wall 
& window (incl. frame) 
& roof 
& cellar ceiling 
& standard
& solar protection \\
\hline
2020 &  0.27&0.19 & 1.00 & 0.22 & 0.29 & 0.50 & 0.20 \\
2000 &  0.54&0.50 & 1.90 & 0.30 & 0.60 & 0.70 & 0.33 \\
1980 &  0.94&0.80 & 3.00 & 0.70 & 0.80 & 0.64 & 0.44 \\
\end{tabular}
\end{table}

\paragraph{Thermal mass} 
The thermal mass of a building influences its heat storage capacity and thermal inertia. The TwinHouses consist of external and ground floor internal walls of 1970 (800 kg/m³) honeycomb bricks, concrete ceilings, and a gypsum board-covered roof truss and internal attic walls. In addition, four concrete pillars on the ground floor support the ceiling structure. Overall, the buildings correspond to a heavy-weight construction, although the attic has a lower thermal mass. Since the TwinHouses serve as the basis for the simulation model, this configuration is defined as the “heavy” case.
The simulation models were further modified to represent medium and light thermal mass variants. For the medium-weight case, the ground-floor internal walls were replaced with drywall constructions, and the four concrete pillars were removed. For the lightweight case, the ground-floor screed was replaced by a dry screed system, and the density of the perforated clay bricks was reduced to represent a construction with larger cavities. In addition, the density of the attic dry screed was reduced.

\paragraph{Window glazing}
We vary the windows' glazing from standard multi-pane products to solar protection glazing to mimic different levels of solar gain. This can similarly be interpreted as different window sizes. The amount of solar short-wave radiation energy penetrating through a glazing is described by the solar heat gain coefficient (SHGC) or the total energy transmission (g-value). Solar protection glazings have lower g-values than standard glazings while maintaining comparable U-values. It must be noted that the U-value of the glazings, which varies with building age, also influences the g-value. So the g-value changes with the usage of solar protection glazing and with building age. The g-values of the chosen glazings are included in Table~\ref{tab:tab_envelope_combined}.

\paragraph{Number of rooms}
The TwinHouses feature seven common rooms, in addition to the unheated rooms, i.e., doorway, corridor, and staircase. To represent different internal room layouts in the dataset, variations with six and five common rooms are created by merging two and three ground-floor rooms in the simulation model. For a six-room house, the kitchen and living room are merged (see Figure~\ref{fig:TH_FloorPlan}). The authors are aware that merging the kitchen and dining room would be more typical, but the TwinHouses' ground-floor plan doesn't allow for this option. For a five-room layout, the kitchen, living, and dining are merged.

\paragraph{Orientation}
The building orientation determines which rooms receive solar gains from incident solar radiation. In the simulation study, the building's orientation is rotated. The rotation is randomly chosen for each case between 0° and 315° in 45° steps. An orientation of 0° corresponds to the as-built TwinHouses configuration, with the living room facing south; 90°, 180°, and 270° correspond to east-, north-, and west-facing living rooms, respectively.

\subsubsection{Occupancy}
\label{sec:occupancy}

This section describes the occupancy-related parameters, including occupancy profiles and the associated comfort conditions defined by temperature setpoints and ventilation rates.

\paragraph{Occupancy profiles}
Representative occupant behavior is modeled using the Occdem software \cite{Flett2021OccDem}, which generates stochastic occupancy profiles describing internal heat gains. Occdem separates these gains into occupant- and appliance-related loads and provides them at room-specific level. The probabilistic formulation ensures variability across both individual profiles and daily patterns throughout the year. As Occdem provides one-year profiles, they are repeated to match the three-year simulation period. The room-specific occupancy schedules are further used to determine the timing of natural ventilation through manual window operation, as described in the paragraph 
\textbf{Ventilation}.
The Occdem profiles are generated at a 10-minute resolution. To match the dataset structure, they are resampled to 30-minute mean values and applied to the 15-minute simulation time step, resulting in occupancy-dependent changes every second time step. For each simulation case, a separate randomly selected occupancy profile is used. Each profile represents a household occupancy of 2 to 4 persons.

\paragraph{Temperature setpoints and night setback}
Temperature setpoints are linked to occupancy to reflect occupant-dependent comfort preferences.
For each building simulation, we sample a base setpoint temperature of 21~\textdegree C with a random offset of $\pm$2.5 K (normal distribution), to represent common comfort ranges. For the sleeping room and children's rooms when no children are generated by Occdem, this base setpoint temperature is reduced by 2$\pm$0.5~K to account for lower temperature requirements in infrequently used spaces. For randomly chosen 70\% of the simulation cases, we consider an additional night setback of 2$\pm$0.5~K of the base setpoint temperature to mimic typical residential nighttime operation. The setback period starts at a randomly selected time between 21:30 and 24:00 and ends between 6:00 and 8:30. For bathroom setpoint temperatures, the base setpoint is increased by 1.5$\pm$1.5 K to reflect higher comfort requirements in the bathroom.
The temperature setpoints and the potential night setback remain constant within each simulation case throughout the three-year period. 

\paragraph{Ventilation}
To capture the influence of air exchange on thermal dynamics, the simulated cases include either mechanical or natural ventilation. For mechanical ventilation, we randomly sample a volume flow rate between 31 and 155~m\textsuperscript{3}/h, based on \cite{GEG2024}. Supply and extract air rooms are defined as shown in Figure~\ref{fig:TH_FloorPlan}. For natural ventilation, window opening times are generated randomly based on the occupancy (see paragraph \textbf{Occupancy profiles}). When occupied, all rooms are ventilated twice a day, typically following the first and last period of occupancy. Building airtightness (uncontrolled leakage of air through the building envelope) is maintained constant across ventilation modes and building ages. For both approaches, all airflows between rooms and between indoor and outdoor environments are modeled using the TRNSYS-TRNflow plugin \cite{trnflow_manual}, a multi-zone airflow model based on COMIS \cite{comis_fundamentals}.

\subsubsection{Building service equipment}

The two variants, BSE1 and BSE2, as described in Section~\ref{ch:RealityLab}, were both implemented in the baseline simulation models. In both configurations, the heat pumps provide space heating and cooling, while BSE2 additionally includes DHW. To represent these operating modes, the simulation model includes three separate heat pump models: one for heating, one for cooling, and one for DHW production. This was done, as heating and cooling require different coefficient of performance (COP) values (describes the ratio between the thermal energy provided by the heat pump and the electrical energy consumption), and the hydraulic model is more stable when separated. Only one of the three heat pump models can operate at a time. The heat pump compressor frequency is PID-controlled to maintain the required outlet temperatures at fixed water flow rates. Electric consumption and thermal energy provided to the water circuits are calculated based on the compressor frequency, the heat pump outlet temperature, and the outdoor air temperature using polynomial functions. The polynomial functions were provided by the heat pump's manufacturer and can not be shared due to proprietary restrictions. 
For the BSE2 setup, a DHW buffer is modeled and connected to both the DHW heat pump and the solar collector. The DHW tapping point, including the corresponding cold-water supply, is realized directly at one of the buffer connections. The water flow rate of the DHW tappings is defined by the individual occupancy profiles.


\subsubsection{Weather}
\label{sec:weather}

The weather strongly influences building thermal dynamics through ambient conditions. We consider five different locations in Central Europe for the simulation study to capture a broad range of operating conditions: Holzkirchen (Germany), Bolzano (Italy), Paris (France), Warsaw (Poland), and Copenhagen (Denmark). The locations were selected from cities in Central Europe with a cold-temperate climate according to \cite{schnieders2020design}. In these regions, buildings require heating during winter and may require cooling during summer. For each location, we generate a three-year weather time series using Meteonorm 8 \cite{meteonorm8}. We create three different typical meteorological years with random seeds and concatenate them to obtain a non-repeating three-year climate file for the TRNSYS simulation. 

\section{Data Records}

The real-world and simulated data are provided separately while following a consistent column structure. The simulated dataset replicates the sensor features of the real-world buildings to ensure comparability and support joint analyses. Both datasets include measurements of heat pump operation, the heating distribution system, the DHW system, weather conditions, and zone-level indoor climate variables. Additional information on the included measurements is provided in Section~\ref{sec:measurements}. Section~\ref{ch:meta_data} further describes the metadata and the file naming convention. The published dataset is available at: \href{https://fordatis.fraunhofer.de/handle/fordatis/486}{https://fordatis.fraunhofer.de/handle/fordatis/486}\cite{ThermBuild} and is organized according to the following folder structure:

\begin{itemize}
    \item Measurement data:
    \begin{itemize}
        \item Raw measurement data (ThermBuild\_measure\_raw.zip): 
        This file contains 15 month of measurement data recorded in the TwinHouses, including data gaps caused by temporary failures of both acquisition systems. Gaps are marked with NaN.
        \item Imputed measurement data (ThermBuild\_measure\_imputed.zip): This file contains the same data as the previous file, with missing values imputed using the kNN method. 
        \item Additional temperature measurements (ThermBuild\_measure\_Temp\_raw.zip): 
        This file includes additional air temperature measurements at four heights (10 cm, 110 cm, 170 cm, and 10 cm below the ceiling) as well as the mean radiant temperature derived from a globe thermometer for all common rooms, which are not included in the simulated data.
    \end{itemize}
    \item Simulation study (ThermBuild\_Sim.zip): This file includes the data of all 958 simulated buildings over the 3-year period.
\end{itemize}

Each ZIP archive contains the time series data of the corresponding buildings in comma-separated value (CSV) format, with one CSV file provided per building.

\newpage
\subsection{File naming convention}
\label{ch:meta_data}

The dataset files follow a strict and descriptive naming convention that encodes all building parameters (metadata) directly in the file name. The file name consists of parameter identifiers (case name), specifying the parameter name, followed by its corresponding value (case id), as summarized in Table~\ref{tab:tab_variants}. 
This structured naming enables direct identification of the files without requiring additional metadata. Table~\ref{tab:tab_variants} also shows the number of variations per parameter in the simulation study for consistency with Section~\ref{ch:Simulation study}. The two real-world datasets from the TwinHouses follow the same naming convention but use the suffix “\_wetReal” to distinguish them from the synthetic weather files and to indicate real Holzkirchen climate.

\begin{table}[t]
\centering
\caption{Overview of the file naming convention. $^{*}$: indicates the parameters of the real-world TwinHouses;
$^{**}$: indicates the value of the realized average air change rate (ACR) of the building in [m\textsuperscript{3}/h].}
\label{tab:tab_variants}
\begin{tabular}{p{1.5cm}p{3cm}p{2.1cm}p{6.5cm}|c}
\hline
\textbf{Case name} & \textbf{Parameter} & \textbf{Case id} & \textbf{id description} & \textbf{Variations} \\
\hline

\_BSE
& Building service equipment
& 1$^{*}$, 2$^{*}$
& iDM iPump; iDM ALM
& 2 \\

\_age
& Building age
& 1$^{*}$, 2, 3
& 2020;
2000;
1980
& 3 \\

\_SoProt
& window glazing type
& 0$^{*}$, 1
& standard glazing; solar protection glazing (according to building age)
& 2 \\

\_mass
& Building thermal mass
& lig, mid, hea$^{*}$
& light; middle; heavy
& 3 \\

\_wet
& Weather
& Hoki, Bolz, .., Real$^{*}$
& Holzkirchen (DE); Bolzano (I); Paris (F); Warsaw (Po); Copenhagen (DK); real& 5 \\

\_size
& Building size
& 080, 100$^{*}$, 120
& 80\%; 100\% (orig size); 120\%
& 3 \\

\_vent$^{**}$
& Ventilation
& (0, 1$^{*}$)
& Natural ventilation; mechanical ventilation (ACR is selected randomly)
& 2, random \\

\_setB
& Night setback
& (0$^{*}$, 1)
& no night setback, incl. night setback
& random \\

\_Use
& Occupancy profile
& (0, ..,1000)
& randomly generated using \cite{Flett2021OccDem}
& random \\

\_rot
& Building orientation 
& [0$^\circ$$^{*}$, ..,315$^\circ$]
& defines south-ward orientation
& random \\

\_roo
& Number of rooms
& 3$^{*}$, 2, 1
& 7 rooms; 6 rooms; 5 rooms
& random \\

\hline
\end{tabular}
\end{table}

\subsection{Included measurements}
\label{sec:measurements}

Table~\ref{tab:tab_reFile} summarizes all measurements included in the dataset. For the thermal systems, the dataset provides flow and return temperatures, storage temperatures at multiple locations, thermal power measurements, and mass flow rates. Supply and return temperatures should only be interpreted when the corresponding mass flow rate is greater than zero, as stagnant water gradually approaches the surrounding air temperature. In addition, the dataset includes electrical power consumption of the heat pump and the corresponding compressor frequency. Auxiliary variables, such as timestamps and heat pump operating modes (heating, cooling, and DHW production), are also provided. At the room level, air temperature and relative humidity are measured at a height of 110 cm. Valve positions and window opening states are normalized between 0 and 1, corresponding to fully closed and fully open, respectively. Finally, the dataset includes the weather measurements.
To ensure a consistent data structure across all buildings, all files share the same column layout. Accordingly, buildings with configuration BSE1 also contain columns related to DHW and solar thermal production, although these entries are set to NaN because the corresponding systems are not present. Similarly, for buildings with fewer than 7 rooms, measurements associated with non-existent rooms are represented by NaN values.

\begin{figure}[htbp]
    \includegraphics[width=1\linewidth]{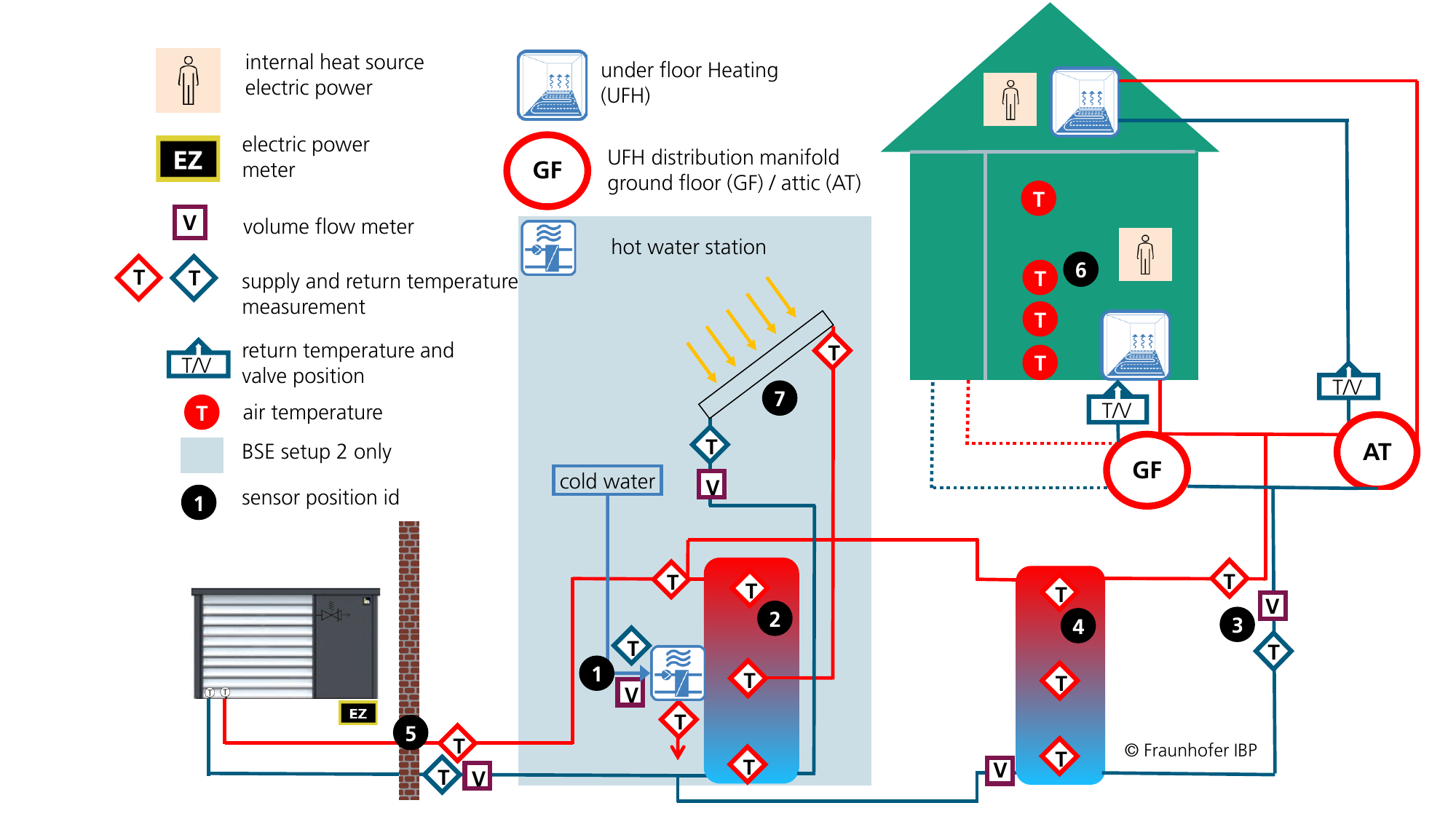}
    \caption{Schematic representation of the instrumentation concept for the heating case.}
    \label{fig:sket_TGA_N2}
\end{figure}

Figure~\ref{fig:sket_TGA_N2} illustrates the measurement concept for the buildings and shows the sensor positions from Table~\ref{tab:tab_reFile}. For the BSE1 system, the grey-shaded area can be neglected. 
For the real-world BSE1 system, only one heat storage sensor is available due to the small tank volume, and the columns \texttt{dist\_Tstore\_top}, \texttt{dist\_Tstore\_mid}, and \texttt{dist\_Tstore\_bott} therefore contain identical values. The simulated BSE1 cases include all three storage sensor positions; however, meaningful stratification cannot be expected because of the limited storage volume.


\begin{table}
\caption{Measurements included in the data files. Measurement names are grouped by system domain: `hp` (heat pump), `dhw` (domestic hot water system), `dist` (heating and cooling distribution system), `solar` (solar thermal collector), `room` (room-specific variables), and `wea` (weather data). Sensor ids refer to the sensor positions shown in Figure~\ref{fig:sket_TGA_N2}. The placeholder `roomID` represents the individual rooms (i.e., child room 1, child room 2, bedroom, bathroom, living room, dining room, or kitchen).}
\label{tab:tab_reFile}
\begin{tabular}{p{3.5cm} c p{9cm} p{2.5cm}} 
\hline
\textbf{Measurement name} & \textbf{Unit} & \textbf{Description} & \textbf{Sensor id}\\ \hline
\texttt{TIME} & h & Simulation time step & \\
\texttt{consecutive\_days} & - & n'th day of the dataset & \\
\texttt{day\_of\_the\_year} & - & day of the year & \\

\texttt{dhw\_Tflow\_store\_tap} & \textdegree C & DHW draw-off temperature & 1\\
\texttt{dhw\_Tstore\_bott} & \textdegree C & Temperature of the DHW tank & 2\\
\texttt{dhw\_Tstore\_mid} & \textdegree C & Temperature of the DHW tank & 2\\
\texttt{dhw\_Tstore\_top} & \textdegree C & Temperature of the DHW tank & 2\\
\texttt{dhw\_tap\_Vset} & kg/h & Setpoint DHW draw-off mass flow & 1\\
\texttt{dhw\_thP\_store\_tap} & kW & Thermal output of DHW tapping & 1\\

\texttt{dist\_Tflow\_store\_ufh} & \textdegree C & Common supply flow temperature of all heating surfaces & 3\\
\texttt{dist\_Tflow\_ufh\_store} & \textdegree C & Return temperature of the entire heating circuit & 3\\
\texttt{dist\_Tstore\_bott} & \textdegree C & Temperature of the heating buffer & 4\\
\texttt{dist\_Tstore\_mid} & \textdegree C & Temperature of the heating buffer & 4\\
\texttt{dist\_Tstore\_top} & \textdegree C & Temperature of the heating buffer & 4\\
\texttt{dist\_Vol} & kg/h & Mass flow in the heating circuit between heating storage and underfloor heating & 3\\
\texttt{dist\_thP} & kW & Thermal output for room heating at the heating buffer outlet & 3\\

\texttt{hp\_elP} & kW & Electrical power consumption of the heat pump & 5\\
\texttt{hp\_rps} & 1/s & Heat pump compressor frequency & 5\\
\texttt{hp\_Tflow\_hp\_store} & \textdegree C & Flow temperature from the heat pump to storage & 5\\
\texttt{hp\_Tflow\_store\_hp} & \textdegree C & Flow temperature from the storage to the heat pump & 5\\
\texttt{hp\_thP} & kW & Thermal power output of the heat pump & 5\\
\texttt{hp\_Vol} & kg/h & Mass flow at the heat pump & 5\\

\texttt{mode\_cool} & bool & System in cooling mode & 5\\
\texttt{mode\_dhw} & bool & Heat pump in DHW production & 5\\
\texttt{mode\_heat} & bool & System in heating mode & 5\\

\texttt{roomID\_ihs} & W & Internal heat source of the room & 6\\
\texttt{roomID\_rh} & \% & Relative room humidity & 6\\
\texttt{roomID\_Tair} & \textdegree C & Air temperature & 6\\
\texttt{roomID\_Tret} & \textdegree C & Return temperature from the room & 3\\
\texttt{roomID\_Tset} & \textdegree C & Room setpoint temperature & 6\\
\texttt{roomID\_valve} & 0-1 & Heating valve control value & 3\\
\texttt{roomID\_win} & 0-1 & Window opening in the room & 6\\

\texttt{solar\_Tflow\_coll\_store} & \textdegree C & Flow temperature from the collector to the DHW buffer & 7\\
\texttt{solar\_Tflow\_store\_coll} & \textdegree C & Flow temperature from the DHW buffer to the collector & 7\\
\texttt{solar\_thP} & kW & Thermal output of the solar thermal collector & 7\\

\texttt{wea\_IbeamHor} & W/m² & Direct solar radiation power on the horizontal plane & wea\\
\texttt{wea\_IdiffHor} & W/m² & Diffuse solar radiation power on the horizontal plane & wea\\
\texttt{wea\_PercentrH} & \% & Relative outside air humidity & wea\\
\texttt{wea\_Tair\_out} & \textdegree C & Outside air temperature & wea\\
\texttt{wea\_vWind} & m/s & Wind speed & wea\\
\texttt{wea\_Wdir} & ° & Wind direction (north=0°; east=90°) & wea\\
\hline
\end{tabular}
\end{table}

\newpage
\section{Technical Validation}

We validated the dataset using several methods. Section~\ref{sec:val_real} describes the sensor validation of the real-world data. For the simulated data, we follow the recommendations of \cite{ZHAI2026116853} and applied two validation approaches: (1) comparison of the simulated data with real-world data, as presented in Section~\ref{ch:Validation of the simulation model}, and (2) validation using “order-of-magnitude” checks, as described in Section~\ref{sec:mag_checks}.

\subsection{Sensor validation for real-world data}
\label{sec:val_real}

Before the measurement campaign started, the sensors followed a defined quality assurance procedure. All room air temperature sensors were placed together in a common radiation shield and compared to identify potential measurement deviations. The water temperature sensors were calibrated in the institute's in-house calibration lab. The installed magneto-inductive mass flow sensors were validated using a calibrated scale by comparing the measured flow rate with the mass of water collected during controlled test runs. The flow meter validation was repeated approximately every 2 months because the hardware occasionally exhibited sensor drift. When drift was detected, the historical measurements were corrected using a linear calibration function, and all flow meters were replaced with manufacturer-calibrated ultrasonic flow meters. The ventilation airflow meters were sent to a certified calibration lab before the start of the measurements.
In addition to calibrating and testing the sensors' quality, two acquisition systems were used for HVAC measurement data, as explained in Section~\ref{ch:methods}. The institute's weather station is subject to a separate quality process. 

\subsection{Comparison of simulation model to TwinHouses}
\label{ch:Validation of the simulation model}

\begin{figure}[t]
    \centering
    \begin{subfigure}[t]{0.6\linewidth}
        \centering
        \includegraphics[height=5cm]{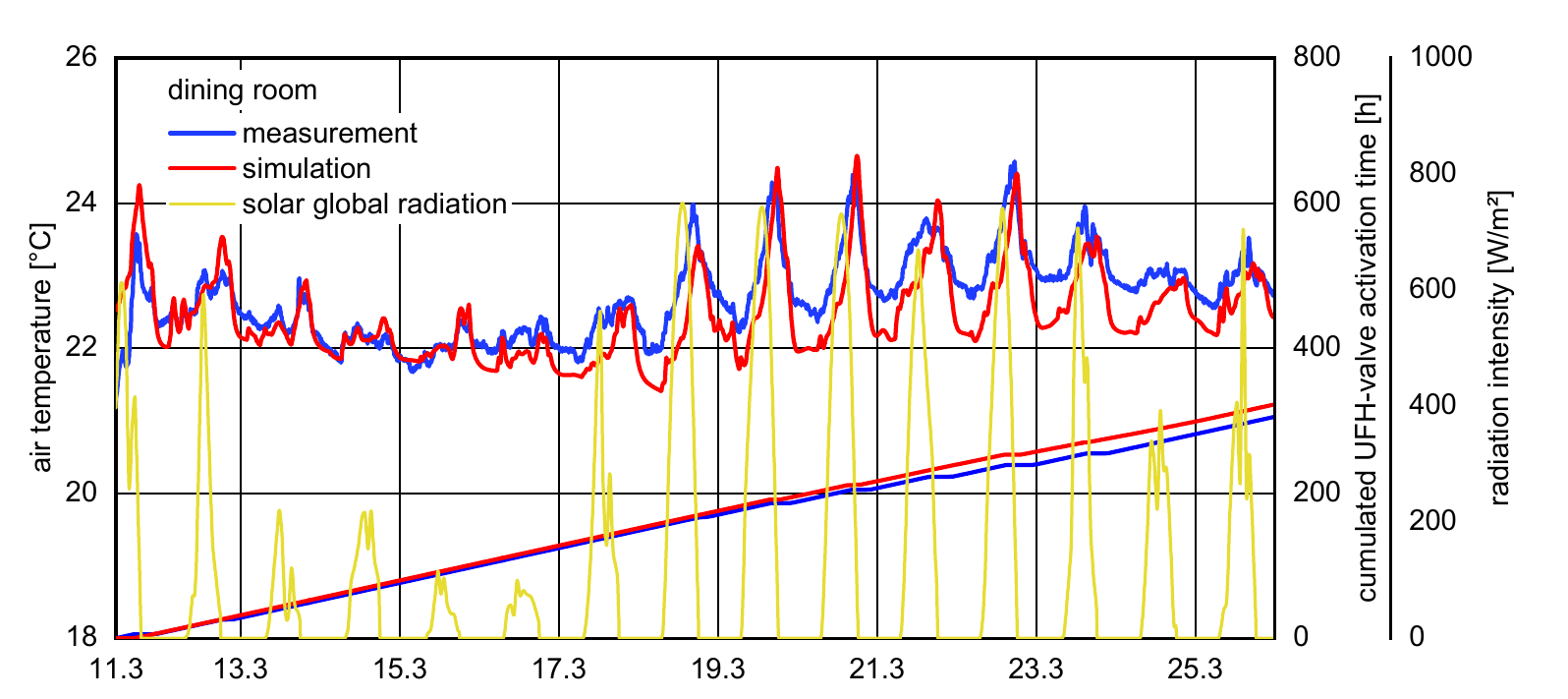}
        \caption{Validation of the O5 dining room.}
        \label{fig:fig_Val_o5Dining_a}
    \end{subfigure}
    \hfill
    \begin{subfigure}[t]{0.39\linewidth}
        \centering
        \includegraphics[height=5cm]{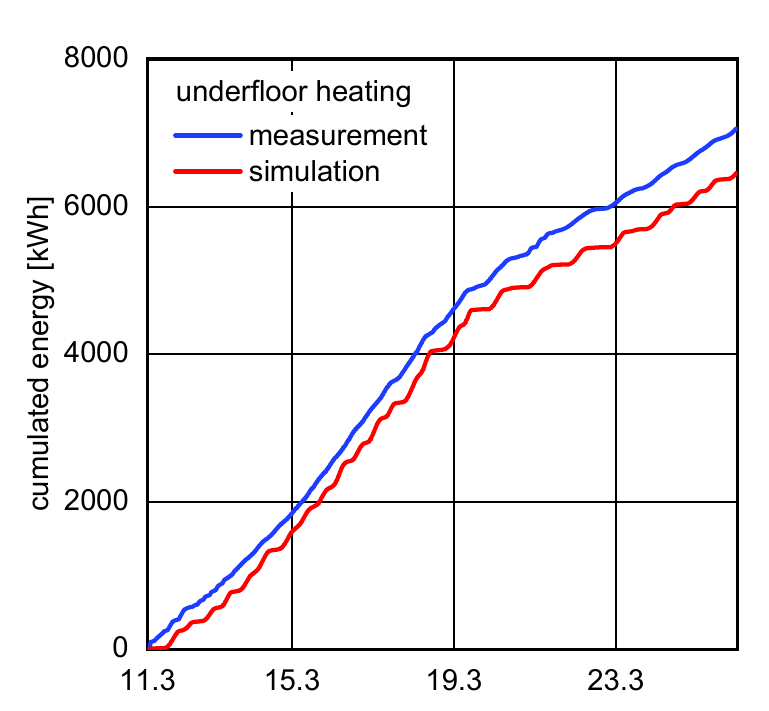}
        \caption{Cumulated heating energy demand of the entire O5 building.}
        \label{fig:fig_Val_o5Dining_b}
    \end{subfigure}
    \caption{Comparison of simulation data to real-world TwinHouse data.}
    \label{fig:fig_Val_o5Dining}
\end{figure}

To validate the TRNSYS simulation model, the measurement data from both TwinHouses were compared with the simulation results. For this comparison, 11--25 March 2025 was chosen because this period included a relatively wide range of outdoor air temperatures and solar radiation intensities and was available at the time of the validation. As an example, Figure~\ref{fig:fig_Val_o5Dining_a} shows the comparison of simulated (red) and measured (blue) room air temperatures and the cumulated activation time for the valve of the underfloor heating for the O5 building's dining room. This validation was additionally performed for all remaining rooms and for the N2 building. Figure~\ref{fig:fig_Val_o5Dining_b} shows the comparison of the simulated and measured heating energy consumption for the O5 building. After the 15-day validation period in March, the accumulated deviation in thermal energy was 8.5\% for the O5 building and 6.9\% for the N2 building. These deviations are within the limits proposed by ASHRAE Guideline 14 \cite{ASHRAE2014}, which allows deviations of up to 10\%. The remaining differences between the TRNSYS model and the measured data can be attributed to several factors: (i) model simplifications regarding thermal properties, control strategies, and boundary conditions, (ii) measurement uncertainties from sensor accuracy and positioning, (iii) differences between on-site micro-climatic conditions and the meteorological input data, and (iv) variations in actual infiltration rates compared to model assumptions.
The deviations could potentially be reduced through parameter optimization. However, this step was intentionally omitted because the objective of the dataset is to represent a broader building stock rather than to create a highly tuned digital representation of the TwinHouses.
In the N2 building, the solar-thermal production was also part of the validation, with a deviation of the cumulated energy gain of 0.4\% at the end of the validation period.


\subsection{Plausibility checks for the simulation study}
\label{sec:mag_checks}

\begin{table}[!b]
    \centering
    \caption{Dependency analysis of building parameter impact based on mean variables of the entire building (x- and y-axes).}
    \label{tab:tab_dep_bild}
    \begin{tabular}{cccc}\hline
         \textbf{x-axis}&  \textbf{y-axis}&  \textbf{color
code}& \textbf{dependency
expected}\\
\hline
         cooling demand&  heat demand
&  age& yes\\
         electricity demand&  heat demand
&  age& yes\\
         cooling demand&  heat demand
&  weather& yes\\
         cooling demand&  heat demand
&  orientation& no\\
         electricity demand&  heat demand
&  BSE& no\\
         electricity demand&  heat demand
&  night setback& no\\
         cooling demand&  heat demand&  window glazing& yes\\
         electricity demand&  compressor frequency&  age& yes\\
         cooling demand&  heat demand&  ventilation& no\\
 supply temperature& compressor frequency& age&no\\
 cooling demand& heat demand& number of rooms&no\\
 electricity demand& supply temperature& age&yes\\
 heat demand& supply temperature& age&yes\\
 cooling demand& heat demand& size&yes\\
 cooling demand& heat demand
& thermal mass&no \\ \hline
    \end{tabular}
\end{table}

\begin{figure}[htbp]
    \centering
    
    \begin{subfigure}[b]{0.32\linewidth}
        \centering
        \includegraphics[width=1.1 \linewidth]{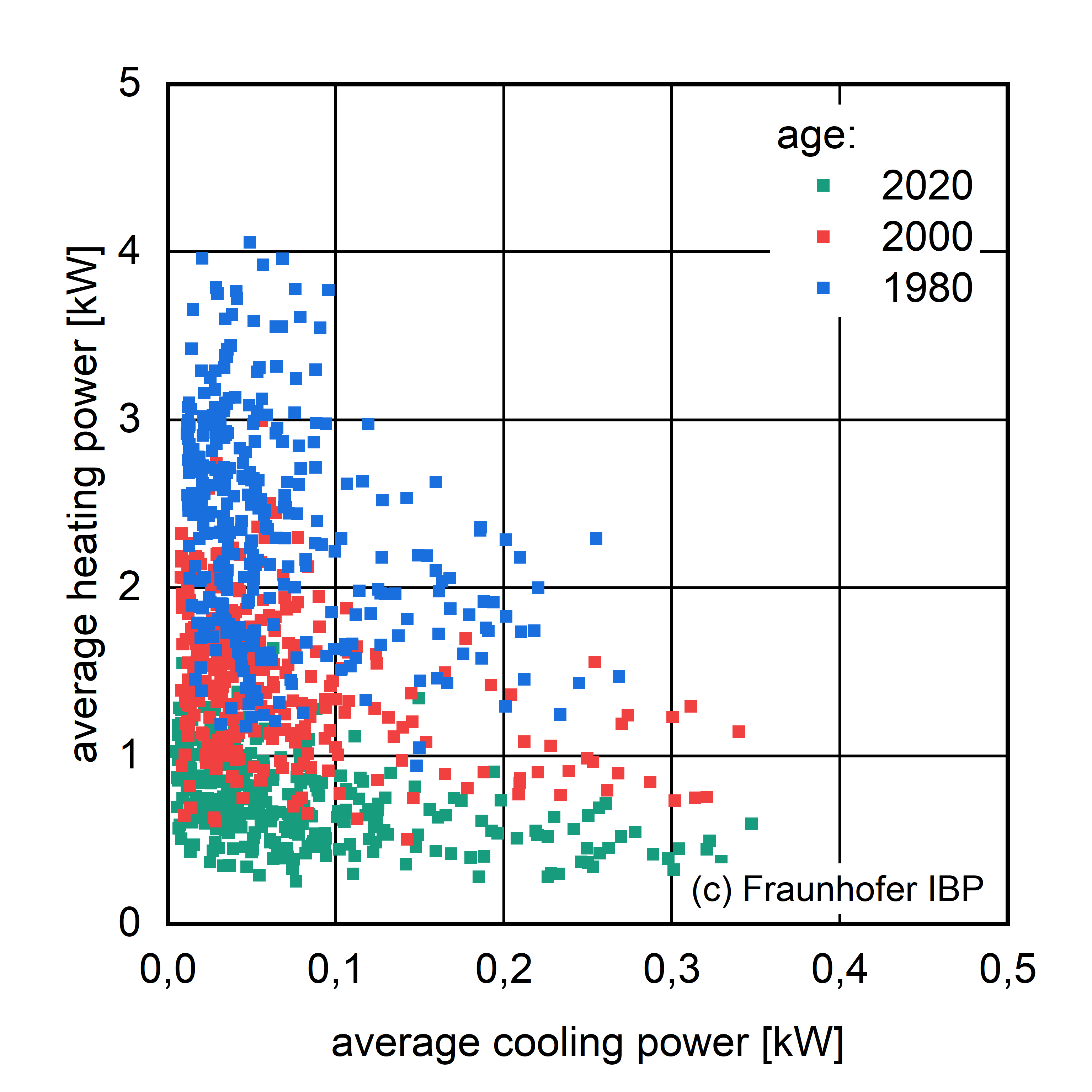}
        \caption{Heating and cooling power vs. building age}
        \label{fig:fig_Plau_Qh_Qc_age}
    \end{subfigure}
    \hfill
    \begin{subfigure}[b]{0.32\linewidth}
        \centering
        \includegraphics[width=1.1 \linewidth]{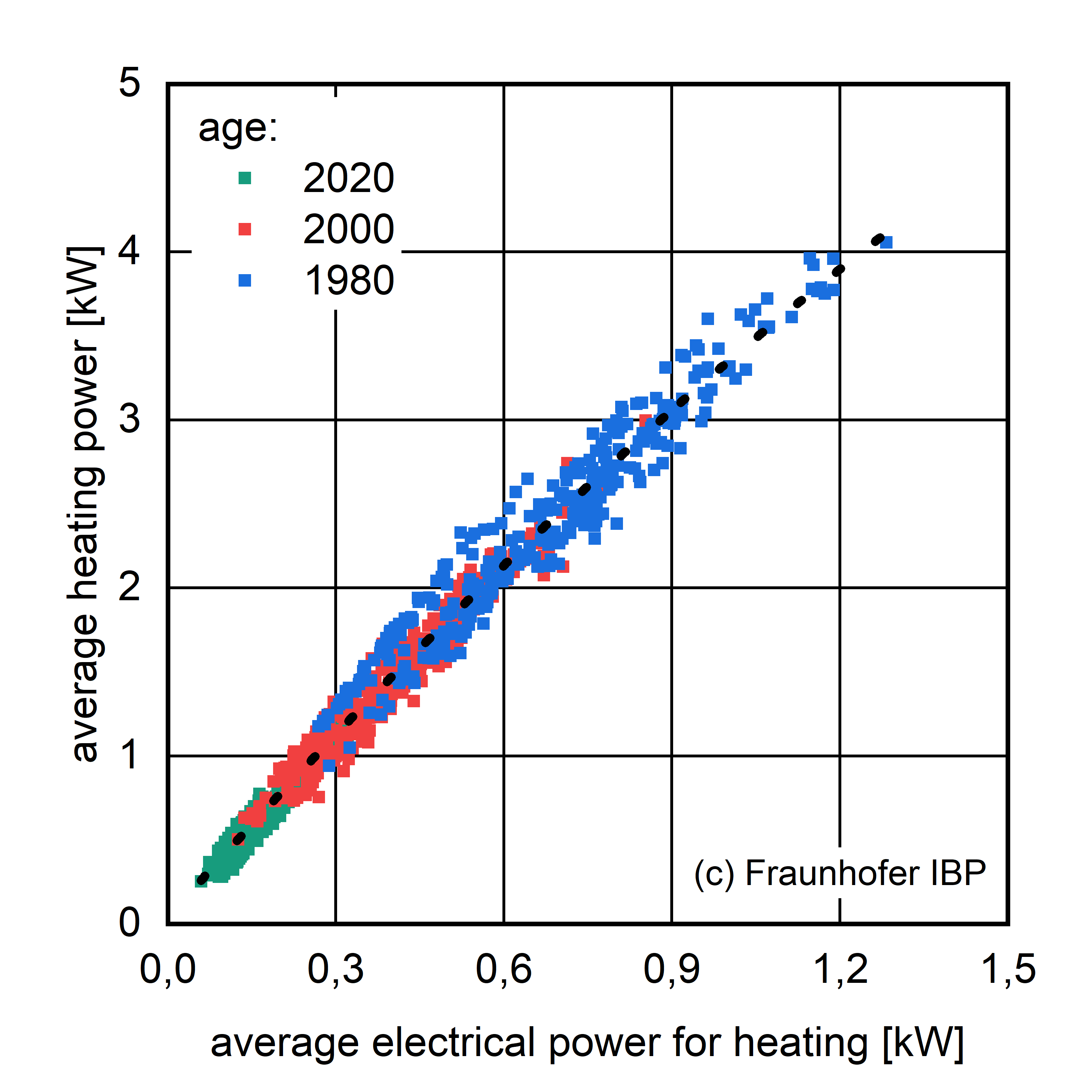}
        \caption{Heating and electrical power vs. building age}
        \label{fig:fig_Plau_Qh_Qe_age}
    \end{subfigure}
    \hfill
    \begin{subfigure}[b]{0.32\linewidth}
        \centering
        \includegraphics[width=1.1 \linewidth]{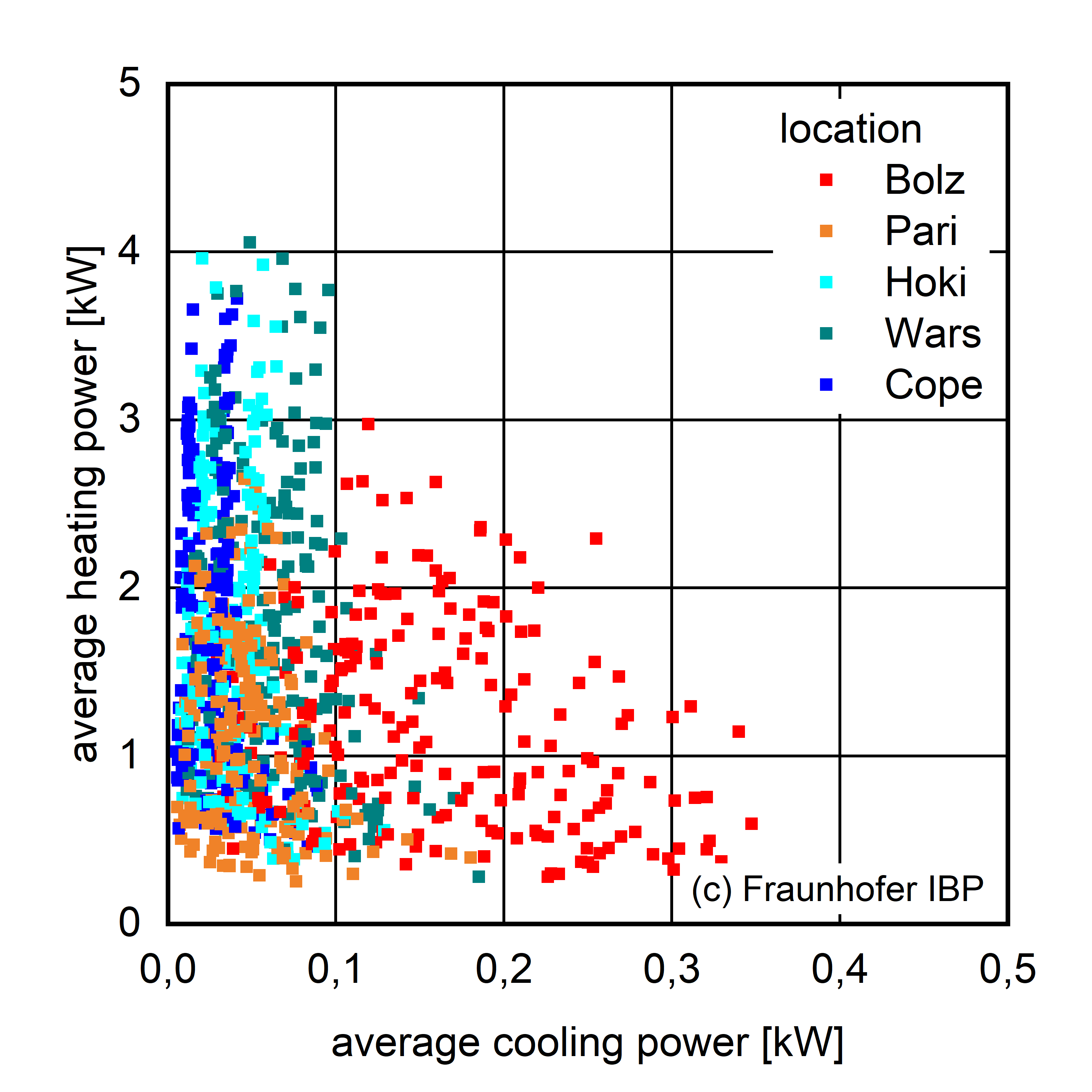}
        \caption{Heating and cooling power vs. weather location}
        \label{fig:fig_Plausi_thP_cthP_location}
    \end{subfigure}
    
    \begin{subfigure}[b]{0.32\linewidth}
        \centering
        \includegraphics[width=1.1 \linewidth]{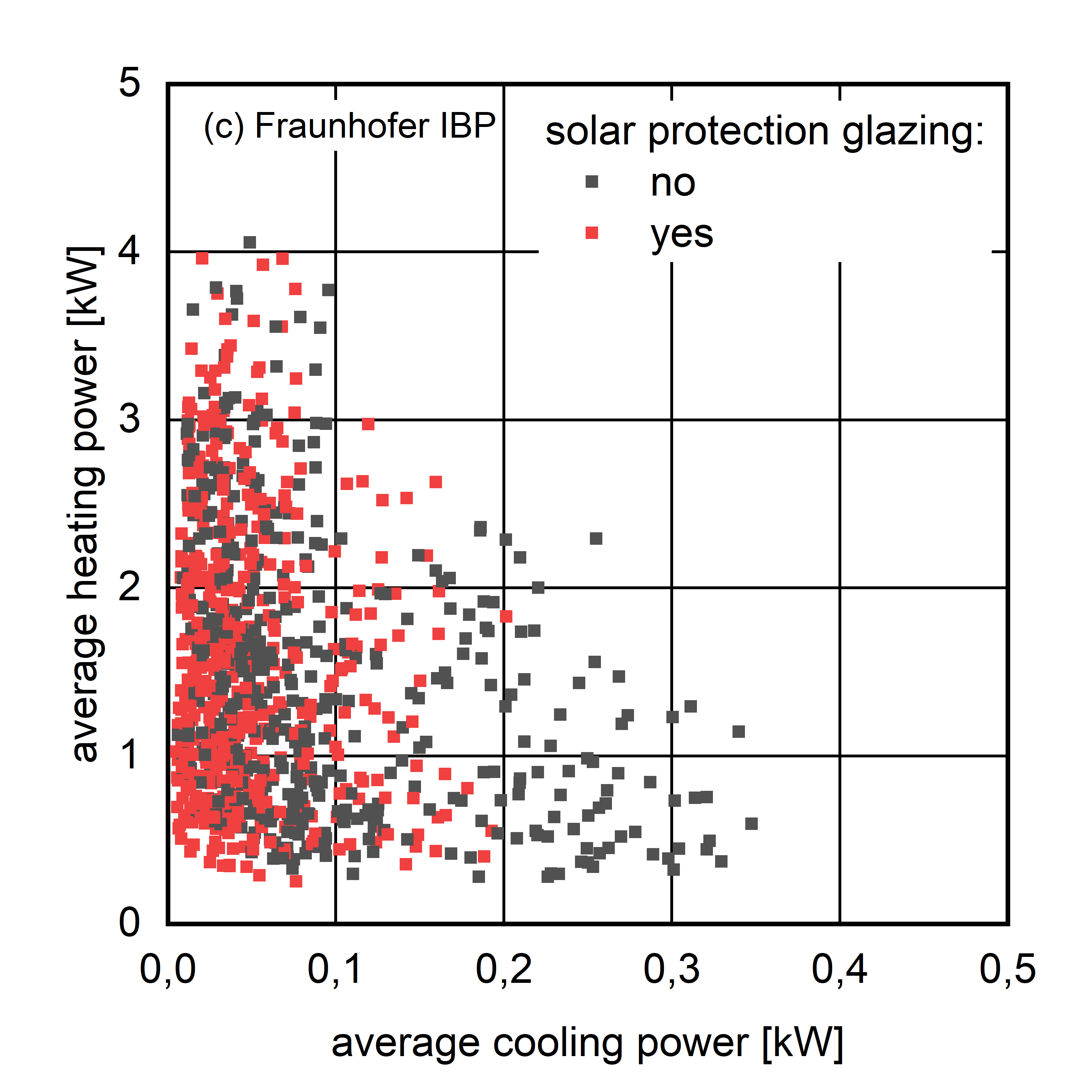}
        \caption{Heating and cooling power vs. glazing type}
        \label{fig:fig_Plausi_thP_cthP_shade}
    \end{subfigure}
    \hfill
    \begin{subfigure}[b]{0.32\linewidth}
        \centering
        \includegraphics[width=1.1 \linewidth]{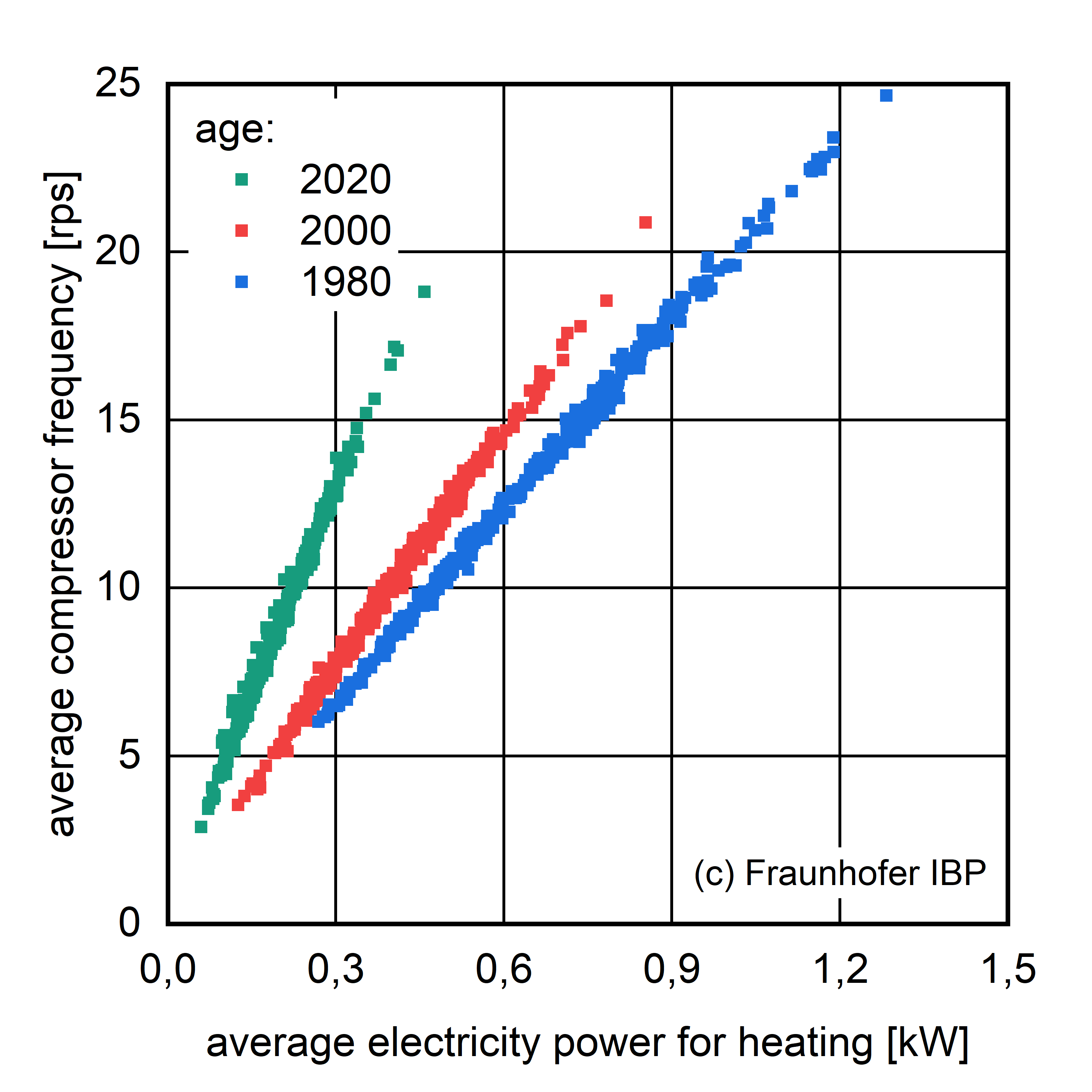}
        \caption{Compressor frequency and electricity consumption vs. building age}
        \label{fig:fig_Plausi_rps_elP_age}
    \end{subfigure}
    \hfill
    \begin{subfigure}[b]{0.32\linewidth}
        \centering
        \includegraphics[width=1.1 \linewidth]{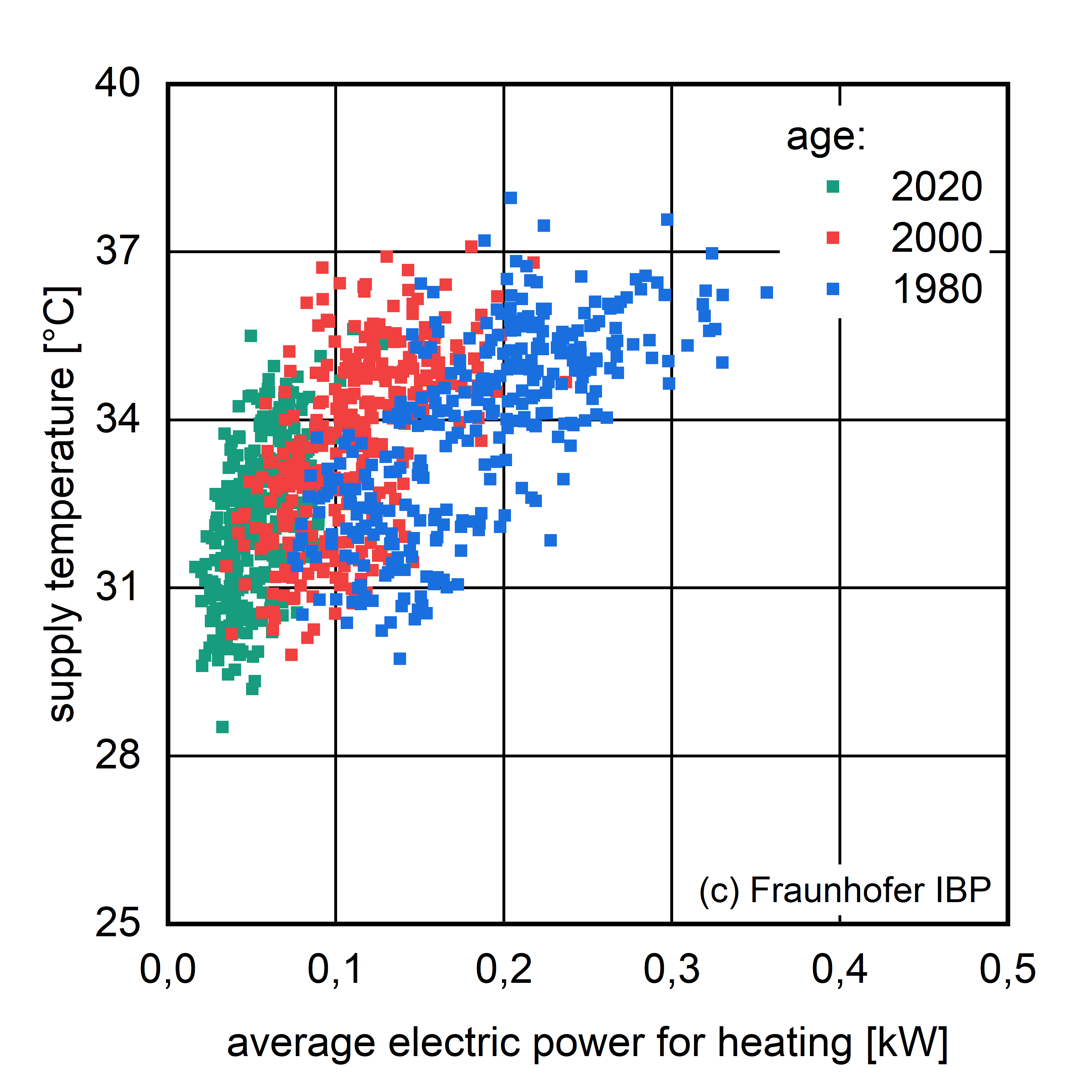}
        \caption{Supply temperature and electricity consumption vs. building age}
        \label{fig:fig_Plausi_elP_Tsup_age}
    \end{subfigure}

    \begin{subfigure}[b]{0.32\linewidth}
        \centering
        \includegraphics[width=1.1 \linewidth]{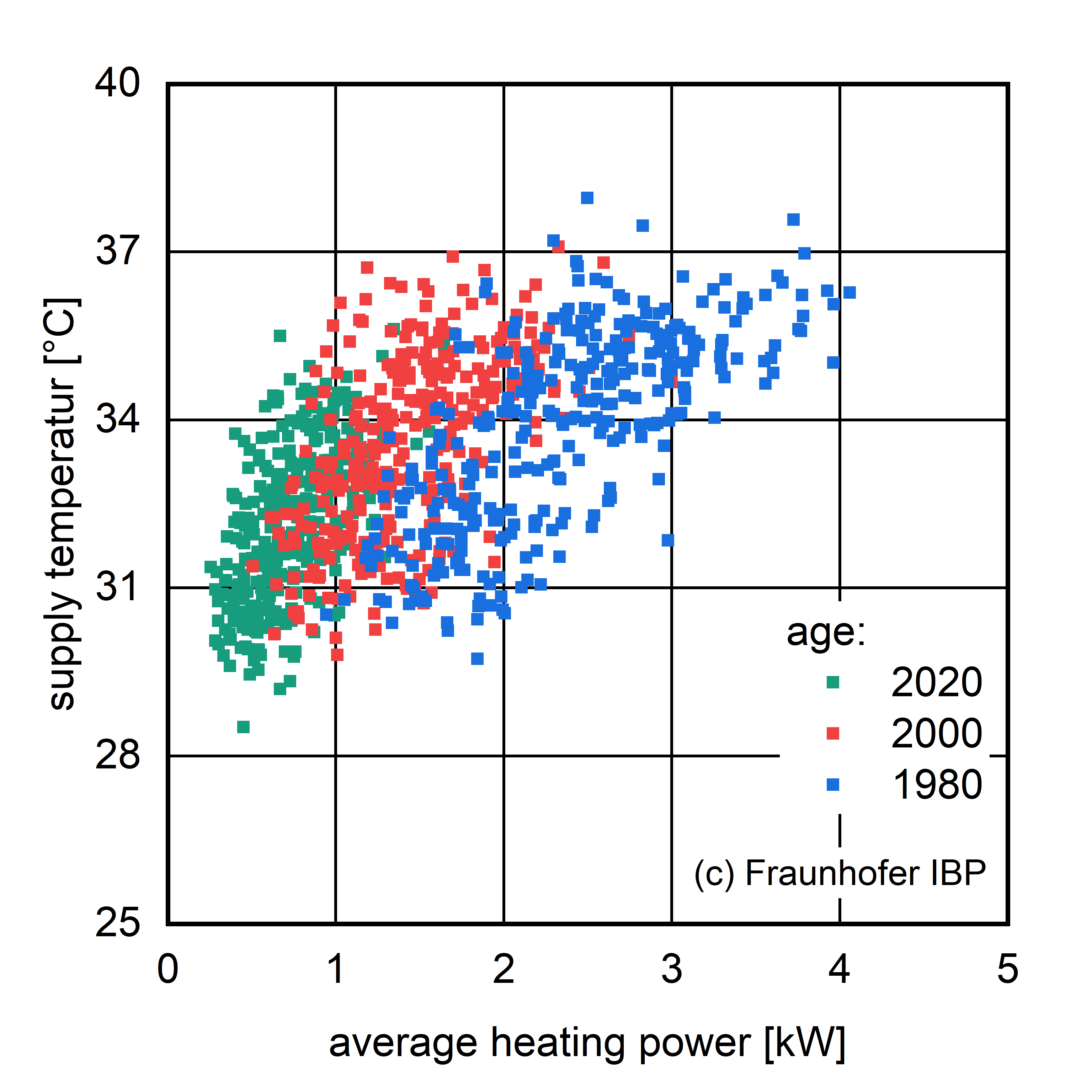}
        \caption{Supply temperature and heating power vs. building age}
        \label{fig:fig_Plausi_thP_Tsup_age}
    \end{subfigure}
    \begin{subfigure}[b]{0.32\linewidth}
        \centering
        \includegraphics[width=1.1 \linewidth]{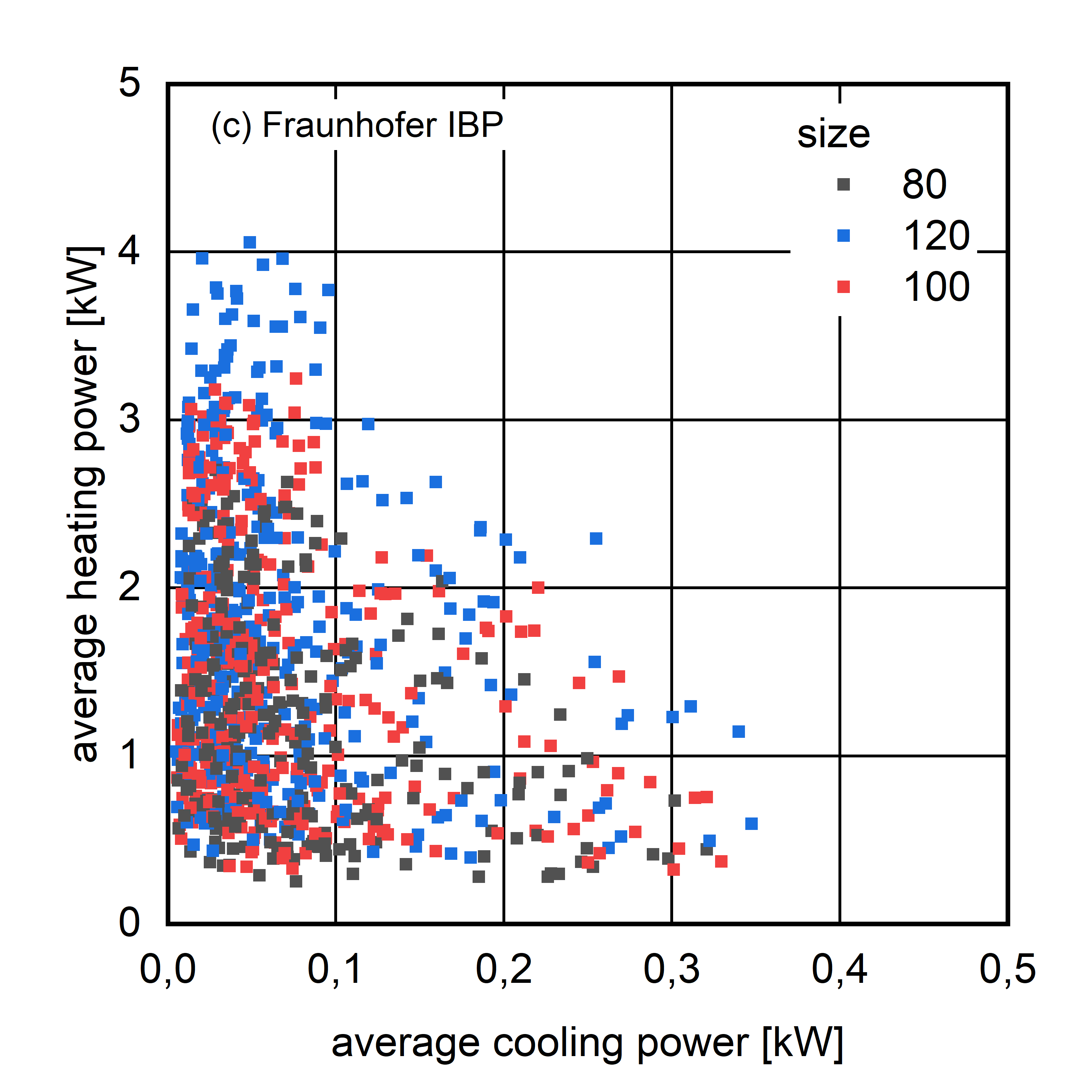}
        \caption{Heating and cooling power vs. building age}
        \label{fig:fig_Plausi_thP_cthP_size}
    \end{subfigure}

    \caption{Scatter plot analysis of simulation results on building level using two measurement variables and one varied parameter.
}
    \label{fig:fig_Plau_combined_entire}
\end{figure}

The final validation step verifies that all parameters varied in the simulation study are correctly implemented in the TRNSYS model and produce the expected influence on the results. To this end, we perform a plausibility analysis by comparing all simulation cases using summary statistics of the three-year time series, as recommended by \cite{ZHAI2026116853}. Table~\ref{tab:tab_dep_bild} summarizes all considered validation cases. 
We focus on the most relevant variables, including those identified as sensitive during the simulation setup, and consider all varied parameters from Figure~\ref{fig:fig_variations}.
Each validation case analyzes the relationship between two variables represented on the x- and y-axes, while the varied parameter is indicated by the color code. For the two variables on the axes, a mean value of the entire building is considered. Later, we will also discuss a room-level validation. 
In Figure~\ref{fig:fig_Plau_combined_entire}, all cases labeled as ``dependency expected'' are plotted (see Table~\ref{tab:tab_dep_bild}).

Figure \ref{fig:fig_Plau_Qh_Qc_age} shows the first example, depicting the dependency between the parameter building age and the variables' average heating and cooling power. It can be seen that the older 1980-buildings (blue) tend to have higher heating demand, while newer buildings (green) have lower heating demand but higher cooling demand. Such behavior is expected. \\
Figure \ref{fig:fig_Plau_Qh_Qe_age} shows the dependency of building age, average heating power, and corresponding electricity consumption. Here, the higher heating demand of older buildings is also observable due to higher heating and electrical power consumption. The curve shows that the thermal-to-electrical ratio bends towards lower coefficients of performance (COP) for buildings with higher heat demand, as can be expected. However, it should be noted that the same heat pump model is used for all age classes (only powers are scaled up according to the heating design load), so there is no distinct COP clustering by age.\\
Figure \ref{fig:fig_Plausi_thP_cthP_location} shows the dependency of average heating and cooling power on the buildings' location. 
Bolzano (dark-gray) has the highest cooling demand because it has the warmest, sunniest climate. Holzkirchen and Warsaw, characterized by the lowest outdoor air temperatures, show the highest heating demands, closely followed by Copenhagen. Paris, with the most average climate among the selected five locations, is not outstanding regarding heating or cooling.\\
Figure \ref{fig:fig_Plausi_thP_cthP_shade} shows the dependency of average heating and cooling power on the type of glazing. It can be seen that cooling loads are reduced with solar protection glazing. No difference in heating demand is observed, as the reduction in solar gains during the darker winter months is relatively small.\\
Figure \ref{fig:fig_Plausi_rps_elP_age} shows the dependency of the average electric power for space heating and the average heat pumps' compressor frequency for the three building ages. First, it is clear that higher heating demands (older buildings) result in a more intensive compressor usage. Second, the difference in building ages results from scaling the heat pumps' capacity to each building's age and its corresponding design heat load. Thus, the 1980 heat pump (blue) delivers 300 W at a compressor frequency of 7 s$^{-1}$, whereas the smaller 2020 unit (green) requires approximately 13 s$^{-1}$ to provide the same thermal power. 
\\
Figure~\ref{fig:fig_Plausi_elP_Tsup_age} shows the average supply temperatures and electrical consumption ratios for different building ages. In addition to the increased electrical consumption observed in older buildings, the average supply temperatures also rise with building age, as higher heating demands require higher design supply temperatures according to the applied heating curves. \\
Figure~\ref{fig:fig_Plausi_thP_Tsup_age} exhibits a similar trend to Figure~\ref{fig:fig_Plausi_elP_Tsup_age}, as the electrical power consumption is replaced by the heating power, which results in a different scaling of the x-axis. \\
Figure~\ref{fig:fig_Plausi_thP_cthP_size} shows the average heating and cooling loads with the building size (80\%, 100\%, 120\%) as the color-coded property. The average heating power depends on the size, since larger buildings have a larger envelope area. For the cooling power, the influence is less clear since not only the window areas, and so the solar gains, scale with the building size, but also the internal wall and ceilings with their thermal mass buffering the solar gains.

\begin{figure}[h]
    \centering
    
    \begin{subfigure}[b]{0.32\linewidth}
        \centering
        \includegraphics[width=1.1 \linewidth]{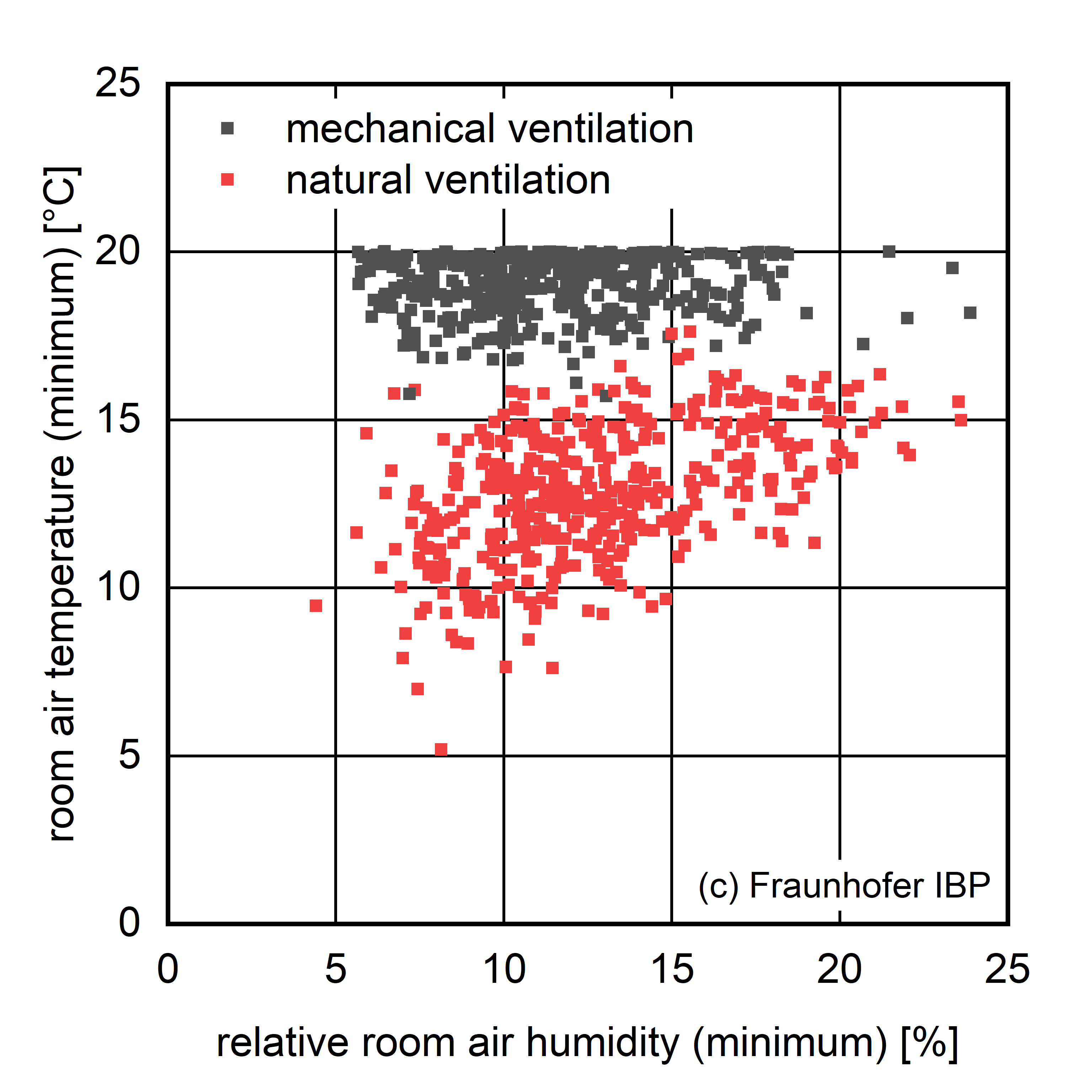}
        \caption{Minimal room air temperature and humidity vs. ventilation type}
        \label{fig:fig_fig_Plau_Tmin_RHmin_vent}
    \end{subfigure}
    \hfill
    \begin{subfigure}[b]{0.32\linewidth}
        \centering
        \includegraphics[width=1.1 \linewidth]{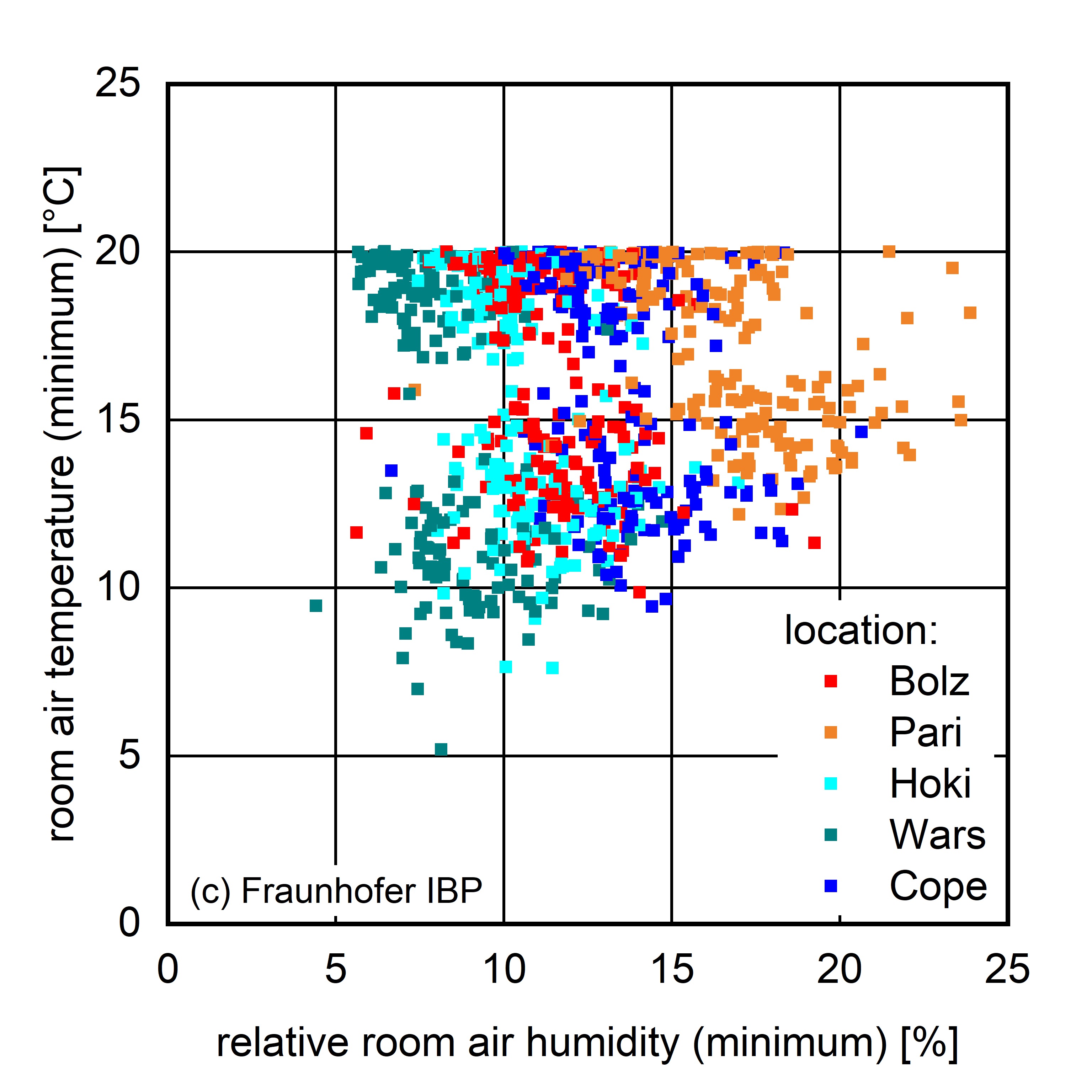}
        \caption{Minimal room air temperature and humidity vs. weather location}
        \label{fig:fig_Plau_Tmin_RHmin_location}
    \end{subfigure}
    \hfill
    \begin{subfigure}[b]{0.32\linewidth}
        \centering
        \includegraphics[width=1.1 \linewidth]{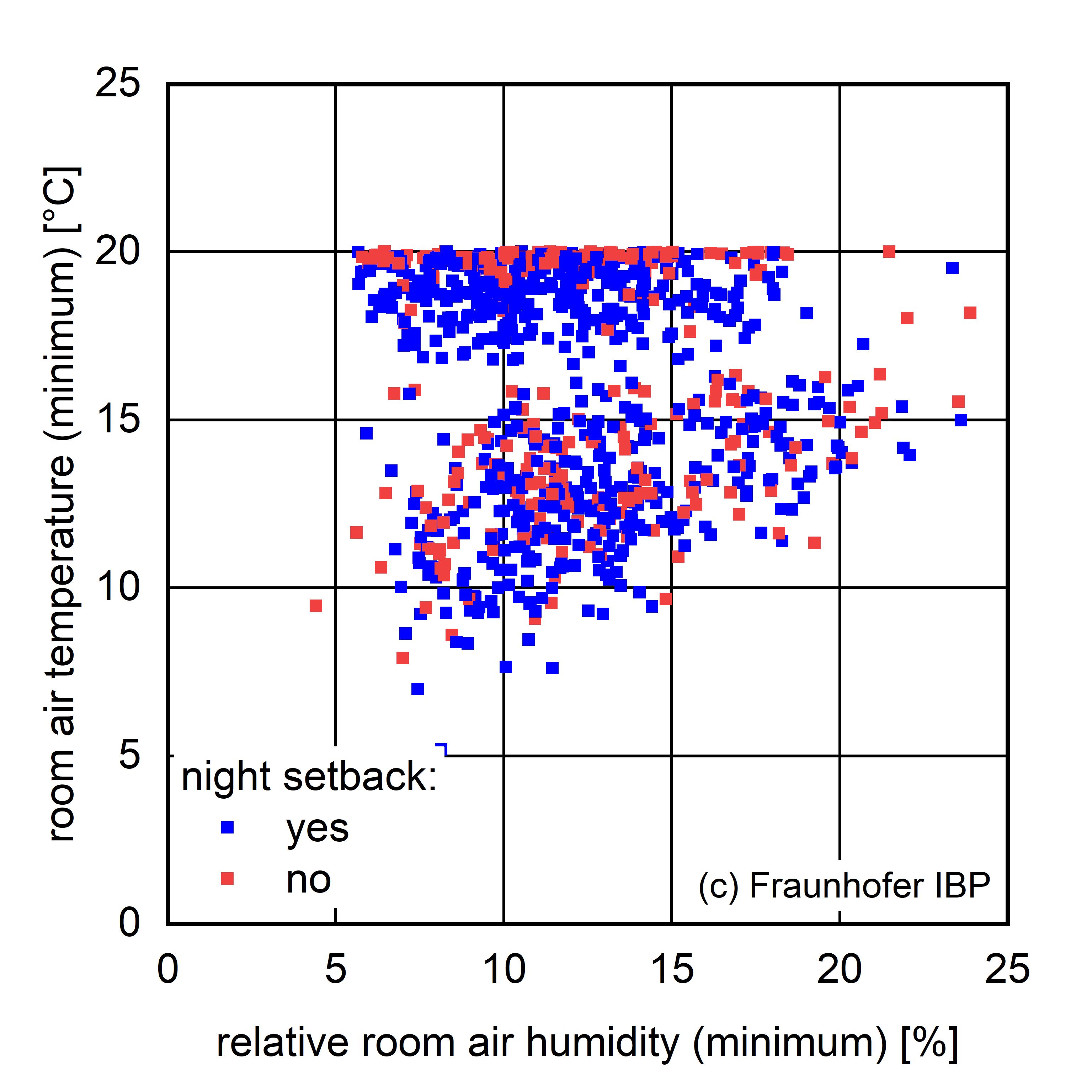}
        \caption{Minimal room air temperature and humidity vs. night setback}
        \label{fig:fig_Plau_Tmin_RHmin_nabs}
    \end{subfigure}
    \caption{Scatter plot analysis of simulation results for the living room.}
    \label{fig:fig_Plau_combined_room}
\end{figure}

Additionally, we perform plausibility checks at room level. The list of room-based checks is documented in Table~\ref{tab:tab_plau_living}. Cases with expected dependency are shown in Figure \ref{fig:fig_Plau_combined_room}. In these three scatter plots, the minimum values for the room air temperature and the relative room air humidity are compared to a color-coded parameter.\\
Figure \ref{fig:fig_fig_Plau_Tmin_RHmin_vent} shows the living room's minimum air temperature and minimum relative air humidity for mechanical and natural ventilation. It can be seen that natural ventilation results in lower minimum temperatures than mechanical ventilation. Mechanical ventilation continuously exchanges the rooms' air at a relatively slow rate. Window openings, on the other hand, lead to brief but intense air exchange with the outside, resulting in very low room air temperatures for a short time. Lower minimum air temperatures correspond to lower minimum air humidities because both result from a higher air change rate during window opening.\\
In Figure \ref{fig:fig_Plau_Tmin_RHmin_location}, the color encodes the location while the x- and y-axes remain unchanged. The two ventilation system groups are still visible. Locations with colder outdoor temperatures in winter and accordingly drier climates show lower minimum room air temperatures and relative humidities.\\ 
Figure~\ref{fig:fig_Plau_Tmin_RHmin_nabs} uses the night setback temperature as the color-coded parameter. The upper block again represents cases with mechanical ventilation and shows that buildings with night setback achieve lower minimum room temperatures because the setpoint temperature is reduced during nighttime operation. In the cases with natural ventilation, the temperature reduction is mainly driven by window opening.

\begin{table}
    \centering
    \caption{Dependency analysis of building parameter impact based on room-level mean and minimum variable values shown on the x- and y-axes.}
    \label{tab:tab_plau_living}
    \begin{tabular}{lcccc} \hline
          \textbf{type}&\textbf{x-axis}&  \textbf{y-axis}&  \textbf{color
code}& \textbf{dependency expected}\\ \hline
          mean&valve time&  air temperature&  size& no\\
          mean&valve time&  air temperature&  window glazing& no\\
          min&rel. humidity&  air temperature&  age& no\\
          min&rel. humidity&  air temperature&  ventilation& yes\\
          min&rel. humidity&  air temperature&  weather& yes\\
 min& rel. humidity& air temperature& thermal mass&no\\
          min&rel. humidity&  air temperature&  night setback& yes\\ \hline
    \end{tabular}
\end{table}


\section{Data Availability}

The ThermBuild dataset is available at \href{https://fordatis.fraunhofer.de/handle/fordatis/486}{https://fordatis.fraunhofer.de/handle/fordatis/486} \cite{ThermBuild}.
The dataset contains real-world measurements from the two TwinHouses over 15 months, including raw measurements, gap-filled measurements generated using the kNN imputation method, and additional room temperature measurement data. Furthermore, the dataset provides 3-year time series data for 958 simulated buildings. For each building time series (real-world and simulated), the corresponding building metadata are encoded in the file name. All data files share a common column structure, and unavailable values are represented as NaN.


\section{Code Availability}
\label{sec:code}
The TRNSYS models, including the associated input data files, are not made publicly available since they contain confidential source code regarding the heat pumps' control. Without detailed information on how the TRNSYS result files are generated, the provided post-processing code cannot be used meaningfully, and other parties cannot repeat the building simulation validation themselves. \\
The Python code for kNN-based imputation of data gaps in the measurements is shown in Listing~\ref{lst:knn_imputation}:

\begin{lstlisting}[
    caption={Script for kNN imputation of missing measurement data.},
    label={lst:knn_imputation}
]
import pandas as pd
import os
from sklearn.impute import KNNImputer

path = "project_path"
filepath = os.path.join(path, "O5_DAMOtl_Messdaten_2026-04-26.csv")
dat_raw = pd.read_csv(filepath)

# identify NaN-only-columns
all_nan_cols = dat_raw.columns[dat_raw.isna().all()].tolist()
valid_cols   = dat_raw.columns[~dat_raw.isna().all()].tolist()

# impute
imputer = KNNImputer(n_neighbors=5)
dat_imputed = imputer.fit_transform(dat_raw[valid_cols])

# result as DataFrame
dat_valid = pd.DataFrame(dat_imputed, columns=valid_cols, index=dat_raw.index)

# add NaN-only columns
dat = pd.concat([dat_valid, dat_raw[all_nan_cols]], axis=1)

# original order
dat = dat[dat_raw.columns]

#save
filepath = os.path.join(path, "O5_DAMOtl_Messdaten_2026-04-26-KNNn5.csv")
dat.to_csv(filepath, index=False, na_rep="NaN")
\end{lstlisting}


\bibliographystyle{unsrtnat}
\bibliography{Literatur}

\end{document}